\documentclass[pre,amsmath,amssymb,aps,twocolumn,nofootinbib]{revtex4-1}
\usepackage{graphicx}
\usepackage{feynmp,color}
\definecolor{labelkey}{cmyk}{.4,.2,0,0}
\usepackage{times}

\newcommand{\Mathematica}[1]{}

\newcommand{\e}{\varepsilon}

\newcommand{\Eq}[1]{Eq.~(\ref{#1})}

\newcommand{\half}{\frac12}
\newcommand{\blue}{\color{black}}
\newcommand{\bea}{\begin{eqnarray}}
\newcommand{\eea}{\end{eqnarray}}
\newcommand{\beq}{\begin{equation}}
\newcommand{\eeq}{\end{equation}}

\newcommand{\be}{\begin{equation}}
\newcommand{\ee}{\end{equation}}

\newcommand{\rme}{\mathrm{e}}
\newcommand{\rmd}{\mathrm{d}}
\newcommand{\nn}{\nonumber}
\renewcommand{\epsilon}{\varepsilon}
\newcommand{\nott}[1]{}
\newcommand{\Fig}[1]{\includegraphics[width=\columnwidth]{#1}} \newcommand{\fig}[2]{\includegraphics[width=#1\columnwidth]{./#2}}

\newlength{\bilderlength}

\newcommand{\cQ}{{\cal Q}}
\newcommand{\cR}{{\cal R}}

\graphicspath{{figures/},{}}

\renewcommand{\paragraph}{\subsubsection*}

\begin{document}

\title{Dynamical Selection of Critical Exponents}
\author{Kay J\"org Wiese}  \affiliation{CNRS-Laboratoire de Physique Th\'eorique de l'Ecole Normale
  Sup\'erieure, 24 rue Lhomond, 75005 Paris, France,}
\affiliation{PSL Research University, 62 bis Rue Gay-Lussac,   75005 Paris, France.}

\begin{abstract}\noindent
In renormalized field theories there are in general one or few fixed points that are accessible by the renormalization-group flow. They can be identified from the fixed-point equations. Exceptionally,  an infinite family of fixed points exists, parameterized by a scaling exponent $\zeta$,   itself function of a non-renormalizing parameter.  Here we report a different scenario with an infinite family of fixed points of which seemingly only one is  chosen  by the renormalization-group flow. 
This dynamical selection takes place in systems with an attractive interaction ${\cal V}(\phi)$, as in   standard $\phi^4$ theory, but where the potential $\cal V$ at large $\phi$ goes to zero, as e.g.\ the attraction by a  defect.
\end{abstract}

\maketitle

\section{Introduction}

The renormalzation group (RG) is a   powerful tool to study critical phenomena of all sorts, be it   liquid-gas transitions, the para  to ferro transition in magnets, or   disordered systems. In many cases one can identify a single or few relevant couplings, and study the RG-flow projected onto this space. An example is the famous $\phi^4$ theory, whose coupling constant $g$ evolves under a change of an infrared scale  $m$.
This framework is well understood. Originally introduced by Wilson \cite{WilsonKogut1974}, it has been treated in many excellent monographs  \cite{BrezinBook,Zinn,CardyBook,Amit}.

Under more general circumstances, a functional RG approach is necessary. Let us start from a microscopic theory with action (energy), in the presence of a source (or background field) $u$,
\be
{\cal S}_u[\phi] = \int\rmd^d x\, \half \left[ \nabla \phi(x)\right]^2 + \frac{m^2}2 \left[ \phi(x)-u\right]^2 + {\cal V}_0(\phi)\ .
\ee  
The partition function  
\be
{\cal Z}[u]:= \int {\cal D}[\phi] \rme^{-{\cal S}_u[\phi]}
\ee
explicitly depends on the source  $u$. 
To 1-loop order the partition function,    evaluated at constant background field $u$, and normalized with its counterpart at ${\cal V}= 0$, reads
\be\label{3}
\ln\! \left( \frac{{\cal Z}[u]}{{\cal Z}_0[u]}\right) = - \int^{\Lambda} \frac{\rmd^d k}{(2\pi)^d} \ln\!\left(1+ \frac{{\cal V}_0''(u)}{k^2+m^2} \right) \ .
\ee
We have explicitly written an  UV cutoff $\Lambda$. 
This equation is at the origin of {\em non-perturbative} renormalization group schemes \cite{HazenfratzHasenfratz1968,WegnerHoughton1973,Polchinski1984,BergesTetradisWetterich2002}, (confusingly) also referred to as {\em exact RG}. 
To leading order, the effective action is $\Gamma(u) = -\ln\! ( {{\cal Z}[u]}/{{\cal Z}_0[u]}) $, and denoting its local part  by ${\cal V}(u)$, we arrive at the following {\em functional} flow equation for the {\em renormalized} potential ${\cal V}(u)$
\be\label{4}
-m \partial_m {\cal V}(u)  = - m \partial_m \int^{\Lambda} \frac{\rmd^d k}{(2\pi)^d} \ln\!\left(1+ \frac{{\cal V}_0''(u)}{k^2+m^2} \right) \ .
\ee
To simplify the treatment, we restrict ourselves to  perturbative RG,   retaining only    terms local in space.  These are the terms of order ${\cal V}''(u)$, and $[{\cal V}''(u)]^2$, leading to 
\begin{figure}
\Fig{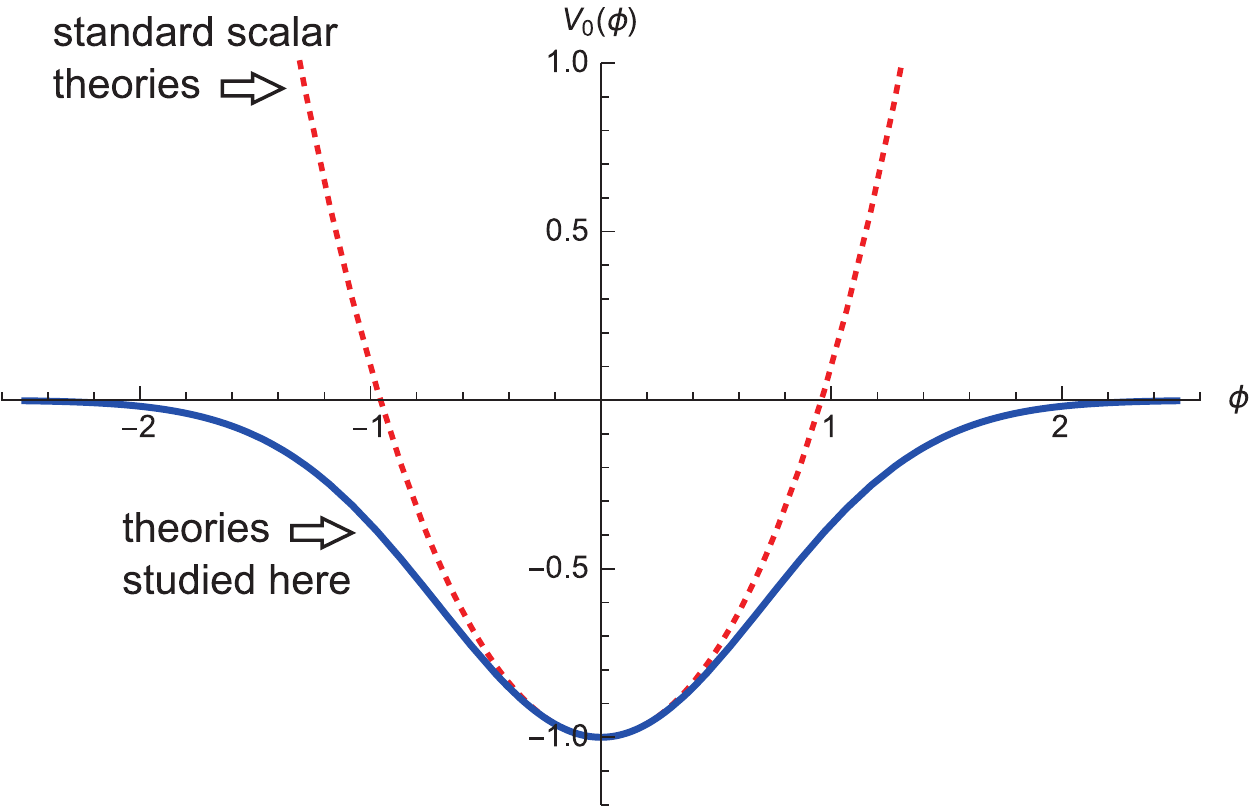}
\caption{The function ${\cal V}_0(\phi)$, for $\phi^4$ theory (top, red, dashed), and a bounded potential (bottom, blue, solid).}
\label{f:VofPhi0}
\end{figure}
\bea
&&\!\!\!-m \partial_m {\cal V}(u) \nn\\
&& = - m \partial_m \int^\Lambda\!\! \frac{\rmd ^d k}{(2\pi)^d} \left[ \frac{{\cal V}''(u)}{k^2+m^2}  \blue - \frac12 \frac{{\cal V}''(u)^2}{(k^2+m^2)^{2}} + ... \right]\ . ~~~~~~~~
\eea
Taking the limit of $\Lambda \to \infty$,   and dropping geometric prefactors, the flow equation becomes 
\be\label{basic}
-m \partial_m {\cal V}(u)  ={\blue m^{d-2}  {{\cal V}''(u)} }  - m^{d-4}  \frac12  {\cal V}''(u)^2  + ...   
\ .
\ee
In the infrared (massless) limit we are interested in, the parameter $m$ becomes small, and the first term can be neglected as compared to the second one, leading to the  simple flow equation
\be
-m \partial_m {\cal V}(u)  = -  m^{d-4}  \frac12  {\cal V}''(u)^2  + ...   \ .
\ee
This equation reproduces the standard RG-equation for the $\phi^4$ theory; indeed, setting 
\be\label{phi4-potential}
{\cal V}(u) = m^{4-d}  \frac{u^4}{72}\, g\ ,
\ee
we arrive with  $\epsilon:=4-d$ at 
\be \label{projection}
-m \partial_m \, g = \epsilon g -  g^2 + ...
\ .
\ee
This is the standard flow equation of $\phi^4$ theory, with fixed point $g_*=\epsilon$. One knows that the potential (\ref{phi4-potential}) at $g=g_*$ is attractive, i.e.\ perturbing it with a perturbation $\phi^{2n}$, $n>2$, the flow will bring it back to its fixed-point form. 

This fixed point, and its treatment with the projected simplified flow equation (\ref{projection}) is relevant in many situations, the most famous being the Ising model. The form of its microscopic potential, which is plotted in figure \ref{f:VofPhi0} (red dashed curve), grows unboundedly for large $\phi$. This is indeed expected for the Ising model, for which the spin, of which $\phi$ is the coarse-grained version, is bounded.

There are, however, situations, where this is not the case. An example is the attraction of a domain wall by a defect. In this situation, one expects that the potential at large $\phi$ vanishes, as  plotted on  figure \ref{f:VofPhi0} (solid blue line). The question to be asked is then: Where does the RG flow lead? 
This is the question addressed in this article. 

As one   sees from figure  \ref{f:VofPhi0},  the {\em bounded} potential ${\cal V}_0$ is negative. In order to deal only with positive quantities, we   set ${\cal V}(u) \equiv - {\cal R}(u)$. The flow equation to be studied is
\be\label{flow-bare}
-m \partial_m {\cal R} (u)  = m^{-\epsilon}  \frac12  {\cal R}''(u)^2  + ...   
\ .
\ee
The potential  ${\cal R} (u)$ appearing in this equation is not dimensionless. Similar to Eq.~(\ref{phi4-potential}), we define a dimensionless function 
\be \label{R-rescaling}
R(u) := m^{-\epsilon +4 \zeta } {\cal R}(u\, m^{-\zeta}  )\ .
\ee
Note that we have allowed for a non-trivial scaling dimension $\zeta$ of the field $u$. The factor of $m^{4 \zeta }$ is necessary to compensate for the dimension of the derivatives. 
With these definitions, the flow equation reads
\be\label{flow-rescaled}
-m \partial_m R (u)  = (\epsilon-4 \zeta ) R(u)+ \zeta u R'(u)+ \frac12R''(u)^2  + ...   \ .
\ee
Note that the RG parameter $m$ appearing in this equation has an intuitive physical interpretation: It is the curvature of the confining parabolic potential, which renders the problem well-defined.

In the remainder of this article, we will show that for generic smooth initial conditions as plotted on figure \ref{f:VofPhi0}: 
\begin{enumerate}
\item [(i)] The flow equations (\ref{flow-bare}) and (\ref{flow-rescaled}) develop a cusp at $u=0$, and a cubic singularity at $u=u_c>0$.
\item [(ii)] Eq.~(\ref{flow-rescaled}) has an infinity of solutions, indexed by $\zeta\in [\frac\epsilon4,\infty]$.
\item [(iii)] The solution   chosen dynamically when starting from smooth initial conditions is $\zeta=\frac\epsilon3$.
\end{enumerate}

\begin{figure}
\Fig{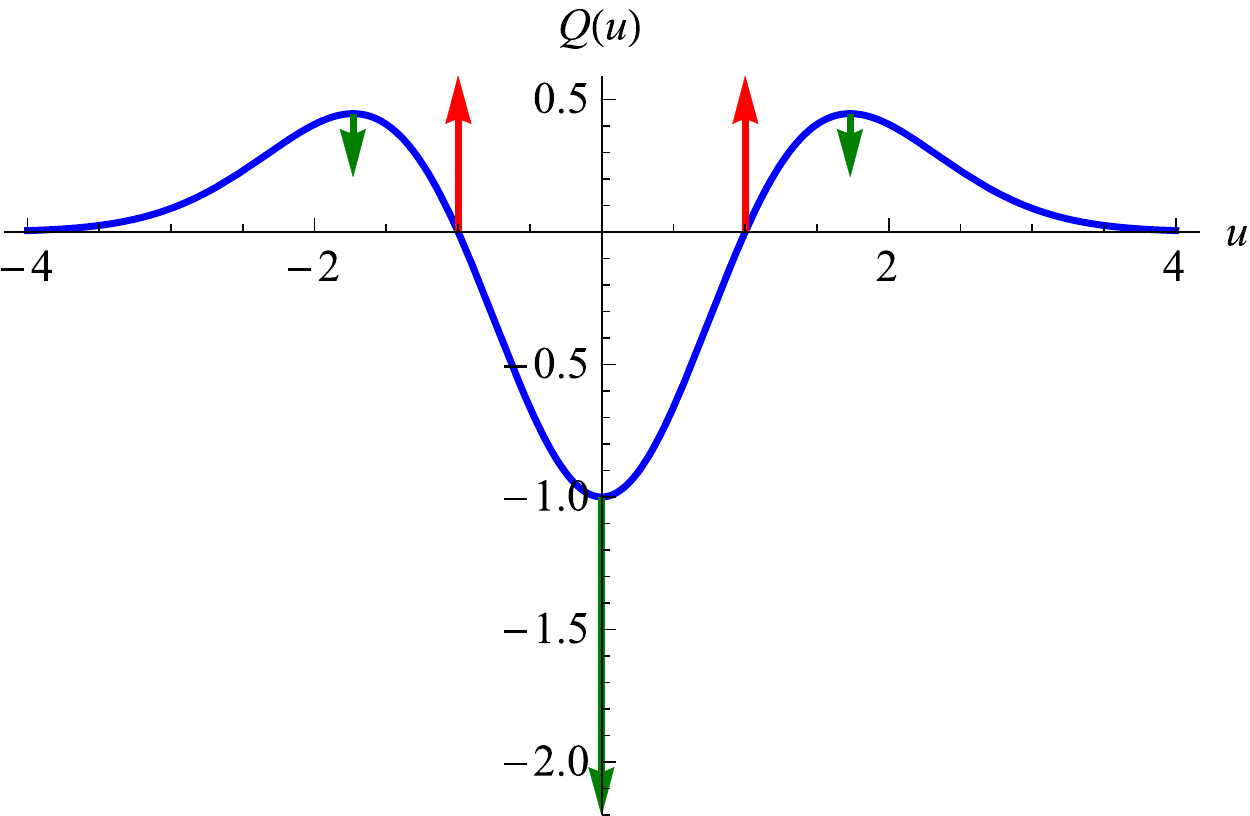}
\caption{The function $\cQ (u) = \cR'' (u^2-1)\rme^{-u^2/2}$, and its flow at the maxima, minima and zeros.}
\label{f:Qflow}
\end{figure}

\section{Some remarks on the literature}
The problem considered here is far from   new, and many articles have been written on the subject, under the denomination of ``wetting''. The latter can be defined as  {\em preferential absorption of one component of a binary liquid by a wall}. Let us summarize the situation:
\begin{enumerate}
\item[(i)] Brezin, Halperin and Leibler \cite{BrezinHalperinLeibler1983a,BrezinHalperinLeibler1983} identified $d=3$ as the critical dimension for wetting. They considered mean-field theory and perturbations around it. This analysis, as several following ones \cite{DSFisherHuse1985,BrezinHalpin-Healy1983}, is based on the  linear term in the functional RG flow equation. 

\item[(ii)] Lipowsky and Fisher \cite{LipowskyFisher1987}, reviewed in Ref.~\cite{ForgasLipowskyNieuwenhuizenInDombGreen}, write down a functional RG equation similar\footnote{The  flow equations (4.14) of \cite{LipowskyFisher1987} and (3.147) of \cite{ForgasLipowskyNieuwenhuizenInDombGreen} contain a log, and not its derivative as our Eq.~(\ref{4}). The reason is that the flow w.r.t.\ the UV cutoff $\Lambda$ is considered. Starting from Eq.~(\ref3) this yields instead of Eq.~(\ref{4}), and up to a multiplicative constant,  $$\Lambda \partial_\Lambda {\cal V}(u)  =    {\Lambda}^d   \ln\!\left(1+ \frac{{\cal V}_0''(u)}{\Lambda^2+m^2} \right) \ .
$$ } to our Eq.~(\ref4). While the analysis in the linear regime follows the same line of reasoning as Refs.~\cite{BrezinHalperinLeibler1983a,BrezinHalperinLeibler1983,DSFisherHuse1985,BrezinHalpin-Healy1983}, they also analyze  the full non-linear flow equations. The  fixed points found all contain a power-law tail, and diverge with a power law at small $u$, or are repulsive. They are thus very different from the fixed points that we will discuss below, and to which a short-ranged initial condition will flow.

\item[(iii)] Another possibility is to consider short-ranged potentials ${\cal V}(u)\simeq \delta(u)$ from the start. This leads to a renormalizable field theory, pioneered by David, Duplantier and Guitter \cite{DDG1,DDG2}, and further studied by several authors \cite{LassigLipowsky1993,Wiese1996a,PinnowWiese2001,PinnowWiese2002a,PinnowWiese2004}. It is this approach that has been successful to tackle the renormalization of self-avoiding manifolds \cite{KardarNelson1987,AronowitzLubensky1987,DDG3,DDG4,WieseDavid1995,DavidWiese1996,WieseDavid1997,DavidWiese1998,DavidWiese2004,WieseHabil}. As the potential ${\cal V}(u)$ has been reduced to a $\delta$-function, any information contained in the shape of ${\cal V}(u)$ is lost.  While this approach is fully non-linear, its domain of applicability is restricted to {\em repulsive potentials}. (We consider attractive potentials.)

\end{enumerate}

\begin{figure}[t]
\Fig{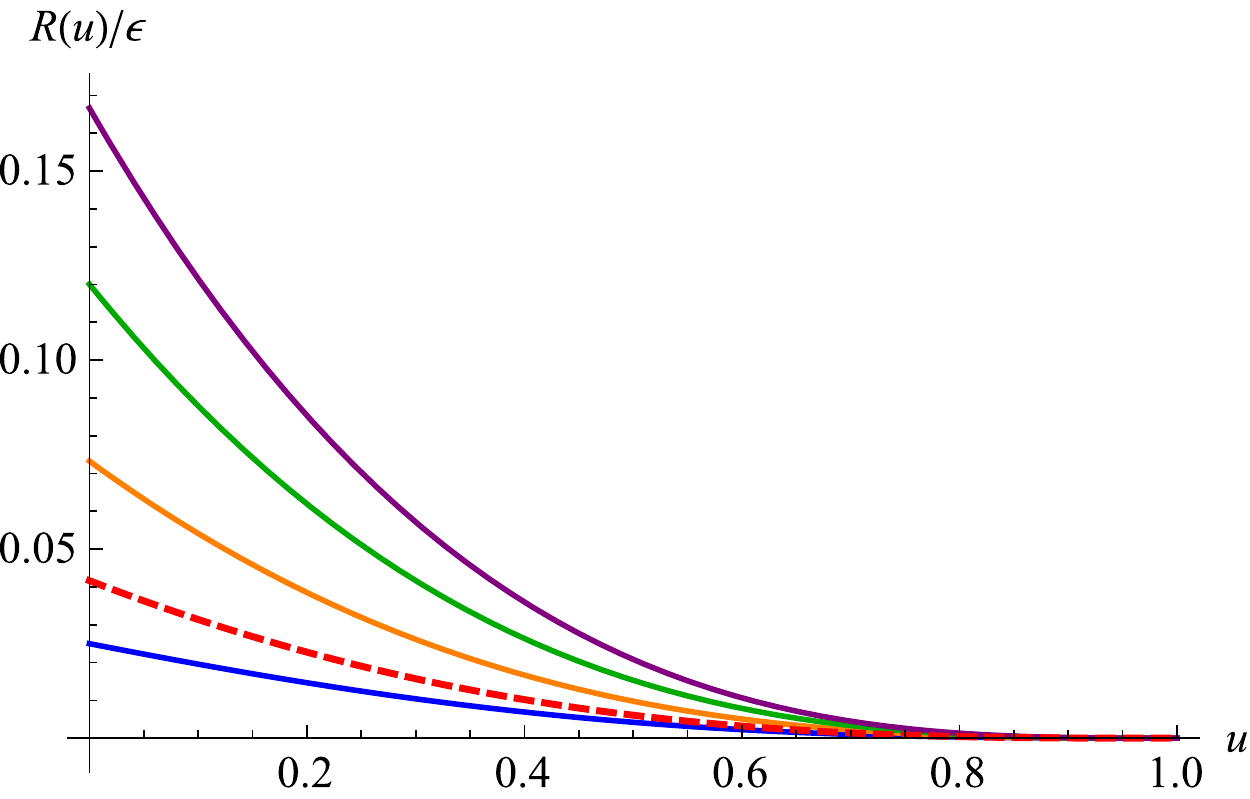}  \caption{From bottom to top: Solutions for $R(u)$ (divided by $\e$, or setting $\e=1$) for  $u_{\rm c}=1$ and  $\zeta=\frac\epsilon 4$ (blue), $\frac\epsilon3$ (red, dashed), $\frac\epsilon2$ (orange), $\frac{3\epsilon}4$ (green) and $\epsilon$ (purple).}
\label{f:someR}
\end{figure}
\section{Cusp formation}
Define 
\be
\ell := \frac1\epsilon \Big(   m ^{-\epsilon} - m_0^{-\epsilon}\Big) \ , 
\ee where $m_0$ is the mass at which we start to integrate the RG equation.
This leads to
\be\label{flow-R-ell}
\partial_\ell {\cal R}(u) =\half {\cal R}''(u)^2\ .
\ee
Further define the curvature function 
\be
{\cal Q}(u):= {\cal R}''(u)\ . 
\ee
It has flow equation
\be\label{65}
\partial_\ell {\cal Q}(u) =   {\cal Q}(u) {\cal Q}''(u)+{\cal Q}'(u)^2\ .
\ee
Let us integrate this flow equation from a microscopic (potentially already coarse-grained) potential 
$
{\cal R}_{\ell=0}(u) = \rme^{-u^2/2}, 
$
i.e.\ 
$
{\cal Q}_{\ell=0}(u) = (u^2-1)\rme^{-u^2/2}. 
$
This function, plotted on Fig.~\ref{f:Qflow},  is negative for $0<u<u_0=1$, and positive for $u>u_0$. It has a zero with linear slope at $u_0=\pm 1$. According to Eq.~(\ref{65})
\be
\partial_\ell {\cal Q}(u_0) =   {\cal Q}'(u_0)^2 > 0\ ,
\ee
and the point $u_0$ will move   towards $0$. 

$Q(u)$ further has a minimum at $u_{\rm min} = 0$ and   maxima at $u_{\rm max}=\pm \sqrt 3$. 
Again according to Eq.~(\ref{65}),  for $u= u_{\rm min}$ or $u=u_{\rm max}$,  
\be
\partial _\ell {\cal Q}(u)=  {\cal Q}(u)  {\cal Q}''(u) <0\ . 
\ee
Thus both the  minimum and the maxima will decrease. 

Since ${\cal Q}(u)$ is a second derivative of an asymptotically vanishing function, it integrates to 0, 
\be
\int_{-\infty}^\infty \rmd u\, {\cal Q}(u) = 0\ .
\ee 
These observations  combined imply that $u_0\to 0$, and that $ {\cal Q}(u) $ must develop a $\delta$-function singularity at $u=0$. 
 Denoting 
\be
{\cal Q}_0 := \lim_{u_0\to 0} \int_{-u_0}^{u_0} \rmd u\, Q(u)\ , 
\ee  
we can decompose the function ${\cal Q}(u)$ into a regular and   a singular part,  
\be
{\cal Q}(u) = {\cal Q}_0 \delta (u) + {\cal Q}_{\rm reg}(u)
\ .
\ee
For the function $\cR(u)$ it implies a cusp at $u=0$, with $\cR'(0^+)={\cal Q}_0/2$. 

\begin{figure}[t]
\Fig{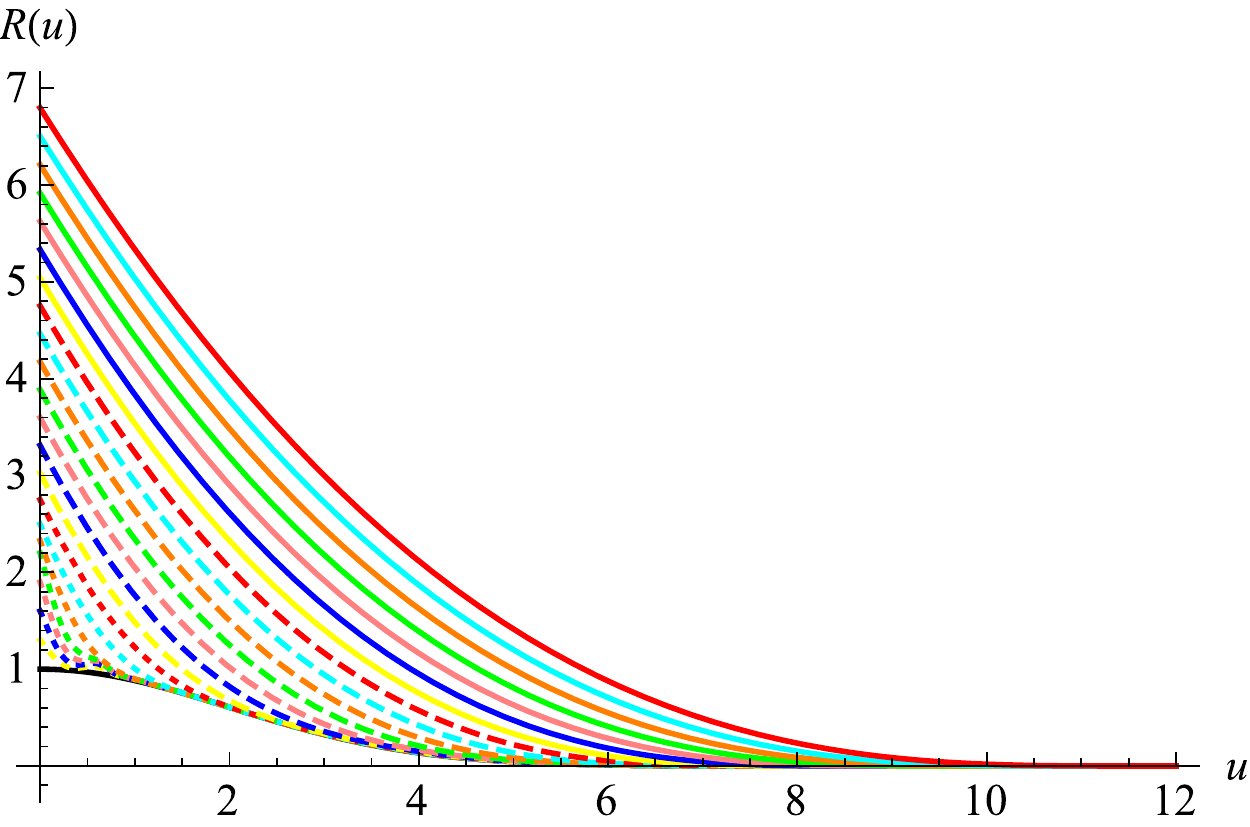}
\caption{Early stages in the development of $R(u)$, starting from the initial configuration $R(u)= \rme^{-u^2/8}$ (black solid line); subsequent configurations have a larger value of $R(0$). The first seven (dotted) configurations are almost at the same time, for the remaining configurations $t_i^{1/3}$ grows linearly. Curves are drawn after $1/\delta t $ iterations, for a total of $ 21/\delta t$ iterations. The last configuration is close to the fixed point with $\zeta=\epsilon/3$.  
}\label{f:R-early-flow}
\end{figure}

\begin{figure*}[t]
\centerline{\Fig{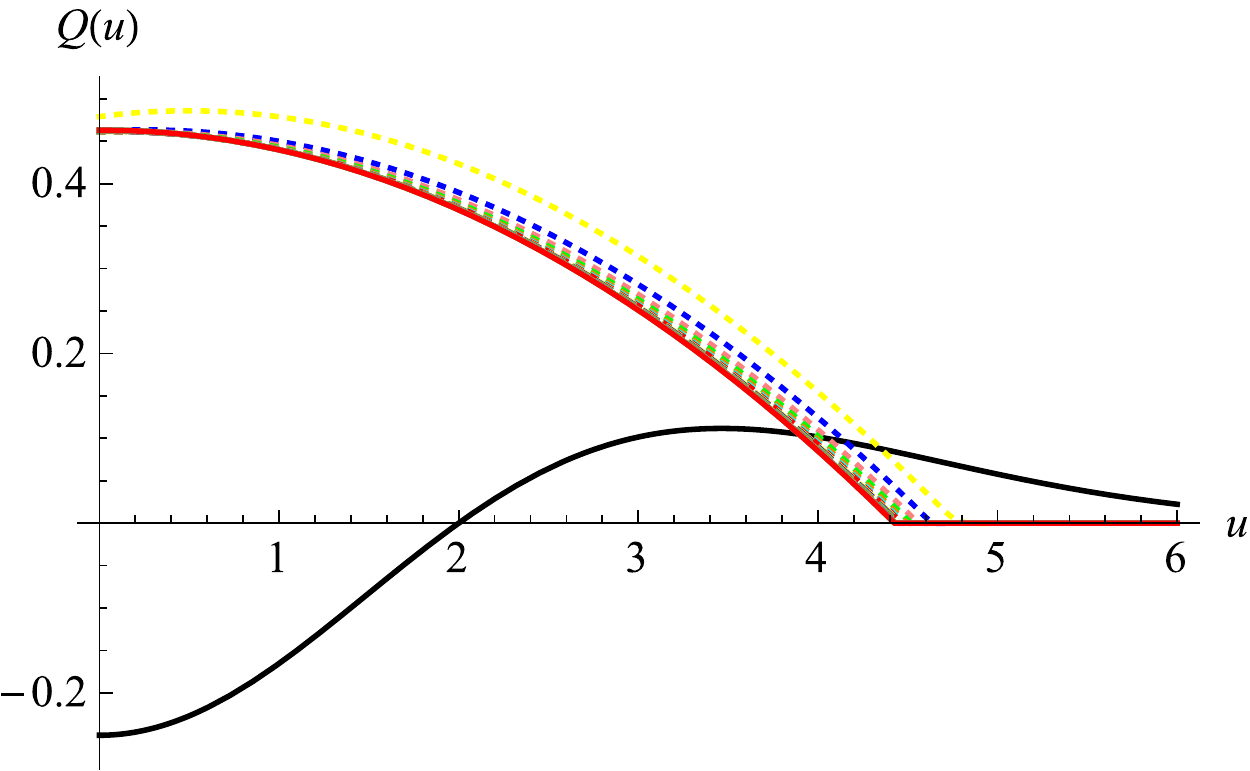}~~~~~~ \Fig{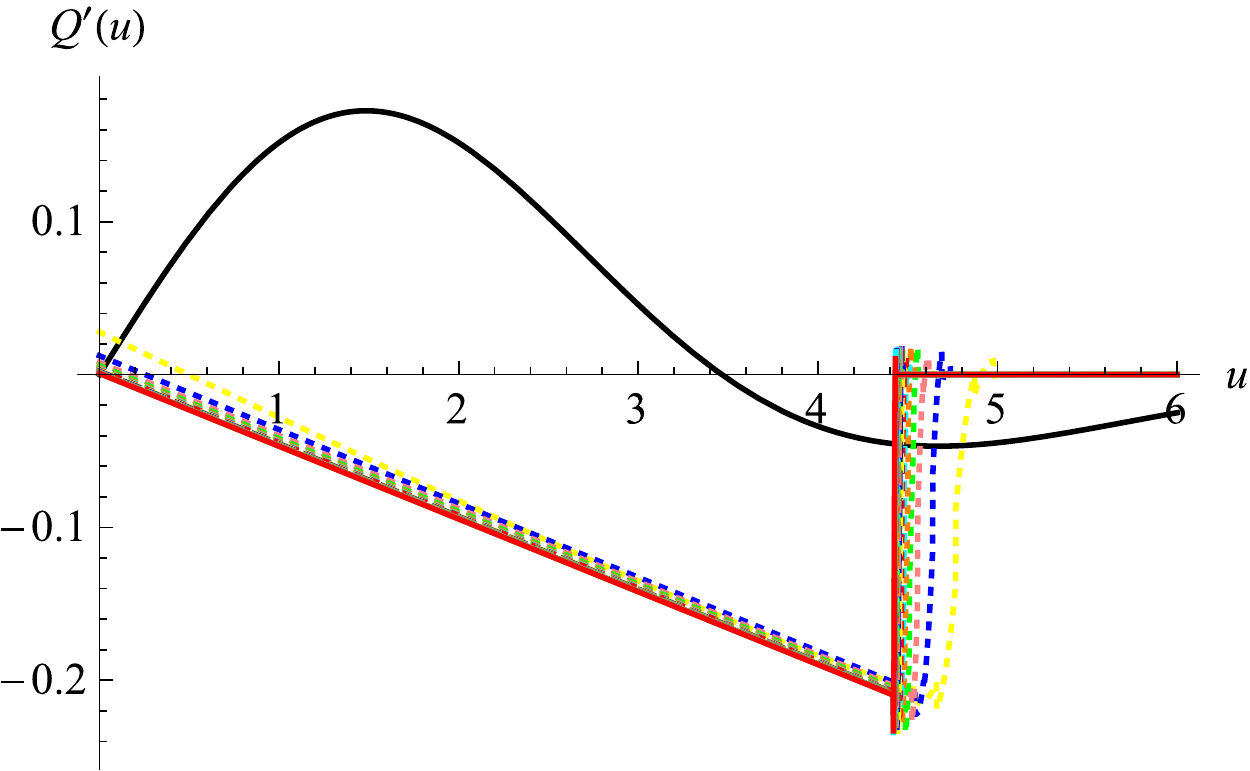}}
\caption{$Q(u)$ and $Q'(u)$ as obtained from the numerical integration of the flow equation,   rescaled with ${  \zeta/\epsilon}=1/3$. 
(Thus the axes are $u \ell^{-1/3}$, as well as  $Q \ell^{1/3}$ and $Q'\ell^{2/3}$.) The initial condition is in black, followed by 21 temporal snapshots, from yellow dashed (the rightmost curves on both plots) to solid red (left-most curve).  One sees how $Q$ converges to an inverted parabola, and $Q'$ to a straight line with ${\cal Q}'(0^+)=0$. The overshoot in the plots for $Q'(u)$ are numerical artifacts.}\label{f:Q}
\end{figure*}

\section{A family of fixed points}
We now search fixed points of Eq.~(\ref{flow-rescaled}). 
 We are   looking for solutions of 
\be\label{FP-eq}
(\epsilon -4 {  \zeta} ) R(u)+ {  \zeta} u R'(u)+ \frac12R''(u)^2 = 0\ .
\ee
We found an infinite family of fixed points, parameterized by the exponent ${  \zeta}$. Some examples are given on Fig.~\ref{f:someR}.
They all have a linear cusp at $u=0$; their Taylor-expansion around $u_{\rm c}=1$ starts with a cubic term,     
\bea\label{R-Taylor}
\!\!\!\! R(u) &=&   \frac{{  \zeta}}{6}   (1-u)^3+\frac{{  \zeta} -\e}{48}  (1-u)^4\nn\\\
&& +\frac{(\e-{  \zeta} ) (3 {  \zeta} -\e) }{1440 {  \zeta} } (1-u)^5\nn\\\
&& +\frac{({  \zeta} -\e) (3 {  \zeta} -2\e)
   (3 { \zeta} -\e) }{17280 {  \zeta} ^2} (1-u)^6 +... 
   \eea
For $u>1$, it vanishes,  $R(u)=0$. For $u<0$, it satisfies $R(u)=R(|u|)$. 
We found the following analytic solutions, valid for $0\le u \le 1$:
\bea \label{R-zeta=1/4}
R_{{  \zeta}=\frac\epsilon 4}(u) &=& \epsilon \left[ \frac{2
   u^{5/2}}{45}-\frac{u^4}{72}-\frac
   {u}{18}+\frac{1}{40} \right] \ ,\\
\label{R-zeta=1/3}
R_{{  \zeta}=\frac\epsilon 3}(u) &=& \epsilon \left[ 
\frac{1}{18} (1-u)^3-\frac{1}{72}
   (1-u)^4 \right] \ ,\\
\label{R-zeta=1}
   R_{{  \zeta}=\epsilon}(u) &=& \frac{\epsilon}{6} (1-u)^3\ .
\eea
The Taylor expansion around $u_{\rm c}=1$, for which the first terms are displayed in Eq.~(\ref{R-Taylor}), converges for $\zeta>\frac \epsilon 4$. 
At $\zeta=\frac\epsilon 4$ the function $R(u)$ develops an additional singularity at $u=0$, and there seems to be no solution\footnote{The convergence radius of the Taylor expansion around 1 is  finite, but smaller than 1 for $\zeta<\frac\epsilon4$.} for $\zeta<\frac\epsilon4$.

\section{Numerical integration of the flow-equations, and fixed-point selection}

We now   integrate numerically the flow equation 
(\ref{flow-R-ell}). 
We solve this equation by discretization in space $u$ and   ``time'' $\ell$. 
Several technical problems need to be considered: First of all, after developing a cusp, the derivative ${\cal R}''(0)$ no longer exists. To integrate the flow equations, we define 
\be
{\cal R}''(0) := \lim_{u\to 0} {\cal R}''(u)\ .
\ee
This limit is obtained by evaluating the discrete second derivative 
\be
{\cal R}''(u) = \frac{{\cal R}(u+\delta u)+ {\cal R}(u-\delta u)-2 {\cal R}(u)}{(\delta u)^2}
\ee
on grid points 2 to 10, and then extrapolating  to the first point with the help of a cubic extrapolation. 

As this procedure seems to be slightly 
  arbitrary,  the following reflection is useful: Instead of making the analysis on the function $\cR(u)$, we can perform it on the function $\cR(|u|)$,  defined on the interval $u \in [0,\infty]$. Derivatives are defined for $u \in ]0,\infty]$. The natural boundary conditions at $u=0$  are   Neumann boundary conditions. This parametrization is   natural at large $N$, see section \ref{s:largeN}.

Second, in order to accelerate the calculations and ensure numerical stability, we iterate
\bea
{\cal R}_{\ell+\delta \ell}(u) &=& {\cal R}_{\ell}(u) + \frac{\delta \ell}2  {\cal R}''(u)^2\ , \\
\delta _\ell &=& \frac{\delta t}{ \max_{u}\{ {\cal R}''(u)^2 \} }\ .
\eea
The parameters chosen were $\delta u = 0.2$, and $ \delta t =   10^{-4}$.
{\blue The result,  plotted as a function of $u \ell^{-1/3}$}, is shown on Fig.~\ref{f:R-early-flow}. The last curve, properly rescaled, is very close to the analytically obtained solution for $\zeta=\frac \e3 $; this is even better seen on the derivatives plotted on figure \ref{f:Q}. That this solution is attained can be checked with several indicators:

First, making the ansatz 
$
\cR_\ell(u) = f(\ell) \, R_{\zeta} \big(g(\ell)u\big) 
$,
  inserting into the flow equation (\ref{flow-R-ell}) and using the fixed point 
condition (\ref{flow-rescaled}) yields
\be
\frac{\epsilon}{\zeta} = {4+\frac{f'(\ell)g(\ell)}{f(\ell)g'(\ell)}}\ . 
\label{zeta-estimate1}
\ee Both $f(\ell)$ and $g(\ell)$ can be measured with high precision, the first from the value of $\cR(0)$, the second from an estimation of the singular point $u_{\rm c}$. 

\begin{figure}[t]
\Fig{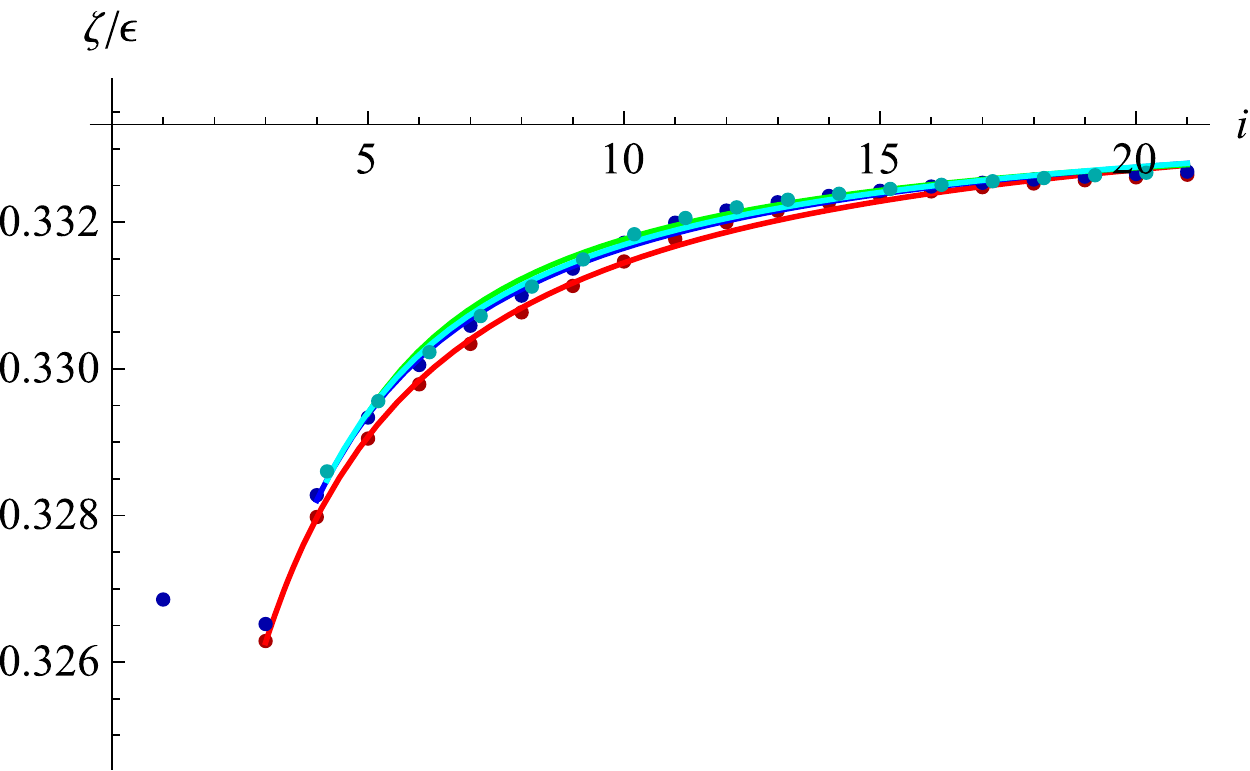}
\caption{Approach to the fixed point for $\zeta=\frac\e 3$, using different indicators, i.e.\ $\zeta$ extracted from $R(0)/R'''(u_{\rm c}) u_{\rm c}^{-3}$ (red), $R(0)/R'(0) u_{\rm c}^{-1}$ (blue), $R''(0)/R(0) u_{\rm c}^2$ (green), and Eq.~(\ref{zeta-estimate1}) (cyan). The index $i$ is the number of iterations in units of $10^5$, and the RG-time is $\ell \simeq 77.4 \,(i-1)^3 $.}
\label{f:zeta-approach}
\end{figure}

Three other functions depending on $\zeta$ can be calculated from the fixed-point solution obtained above, and inverted to give $\zeta$:   
\bea \label{estimator1}
e_{1}(\zeta) &=& \frac{\cR(0)}{\cR'''(u_{\rm c}) u_{\rm c}^{3}}\ ,\\
\label{estimator2}
e_{2}(\zeta) &=& \frac{\cR(0)}{\cR'(0) u_{\rm c}}\ ,\\
\label{estimator3}
e_{3}(\zeta)  &=&  \frac{\cR''(0) u_{\rm c}^2}{\cR(0) }\ .
\eea
The results from all four indicators   are plotted on figure \ref{f:zeta-approach}. They consistently show that the numerically obtained solution converges towards the solution with $\zeta=\frac\epsilon 3$.

All fixed-point  solutions found above have, for $u$ close to $u_{\rm c}$, the form 
\be\label{35}
\cR_\ell(u) \simeq {\cal A}\, [u_{\rm c}(\ell)-u]^\alpha \Theta \big(u<u_{\rm c}(\ell)\big)\ , 
\ee
with $\alpha=3$. This is understood as follows: Making the ansatz (\ref{35}), and imposing the flow equation  
$
\partial _\ell R\big(u,u_{\rm c}(\ell)\big) \stackrel ! = \half \left[ \partial_u^2 R\big(u,u_{\rm c}(\ell)\big) \right]^2
$
implies 
\beq
{\cal A} \alpha \left[ u_{\rm c}(\ell){-}u\right]^{\alpha-1} \partial_\ell u_{\rm c}(\ell) \stackrel ! = \half \left[{\cal A} \alpha(\alpha{-}1)\right]^2 \left[ u_{\rm c}(\ell){-}u\right]^{2\alpha-4}
.
\eeq
The same power-law on both sides is achieved for $\alpha=3$,  which also gives the front velocity as
\be
\partial_\ell \,u_{\rm c}(\ell) = 6 {\cal A}\ .
\ee
We also remark that given $u_{\rm c}$, the amplitude  $\cR(0) $ is {\em not} fixed by the flow equation (\ref{flow-R-ell}), even though it is fixed for the   solutions of Eq.\ (\ref{FP-eq}).  Indeed, changing $\cR(u)\to \kappa \cR(u)$, this can be absorbed into a change of $\ell \to \kappa \ell$.  This is the reason why in Eqs.~(\ref{estimator1}) to (\ref{estimator3}) all ratios are invariant under both a rescaling of $\cR$ and $u$.

\begin{figure*}
\centerline{\fig{1}{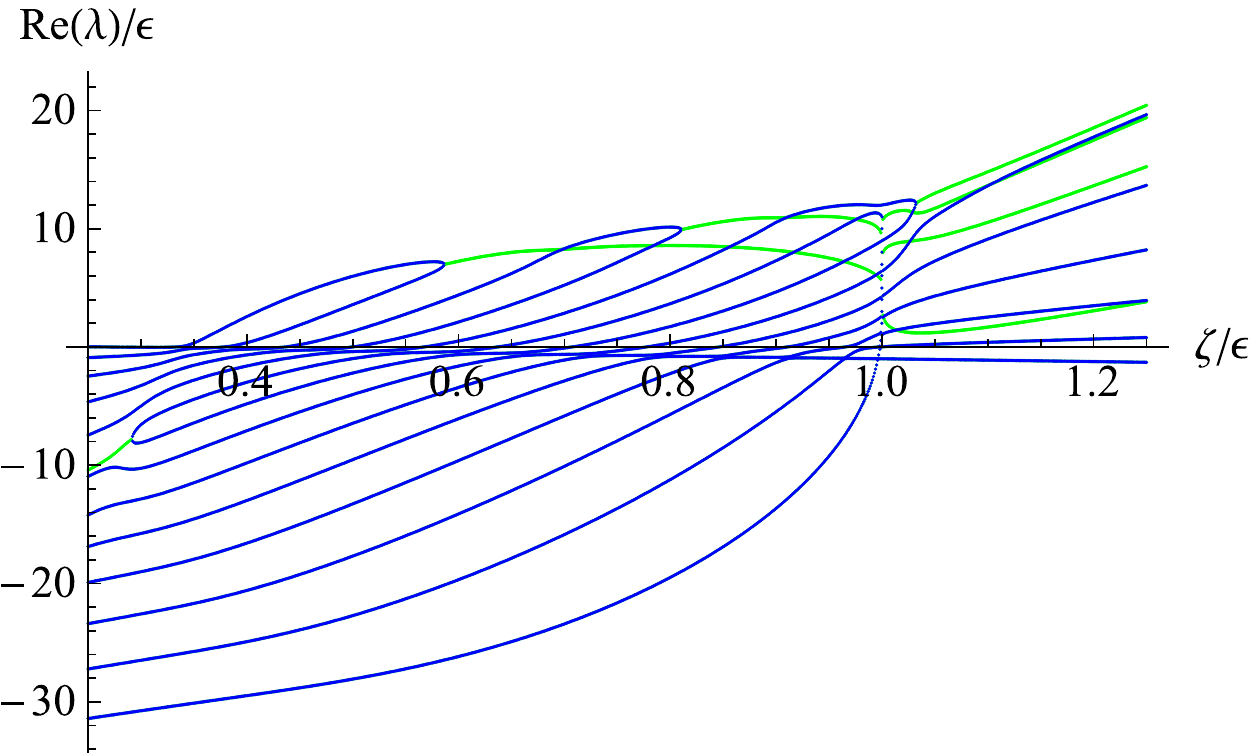}~~~~~~
\fig{1}{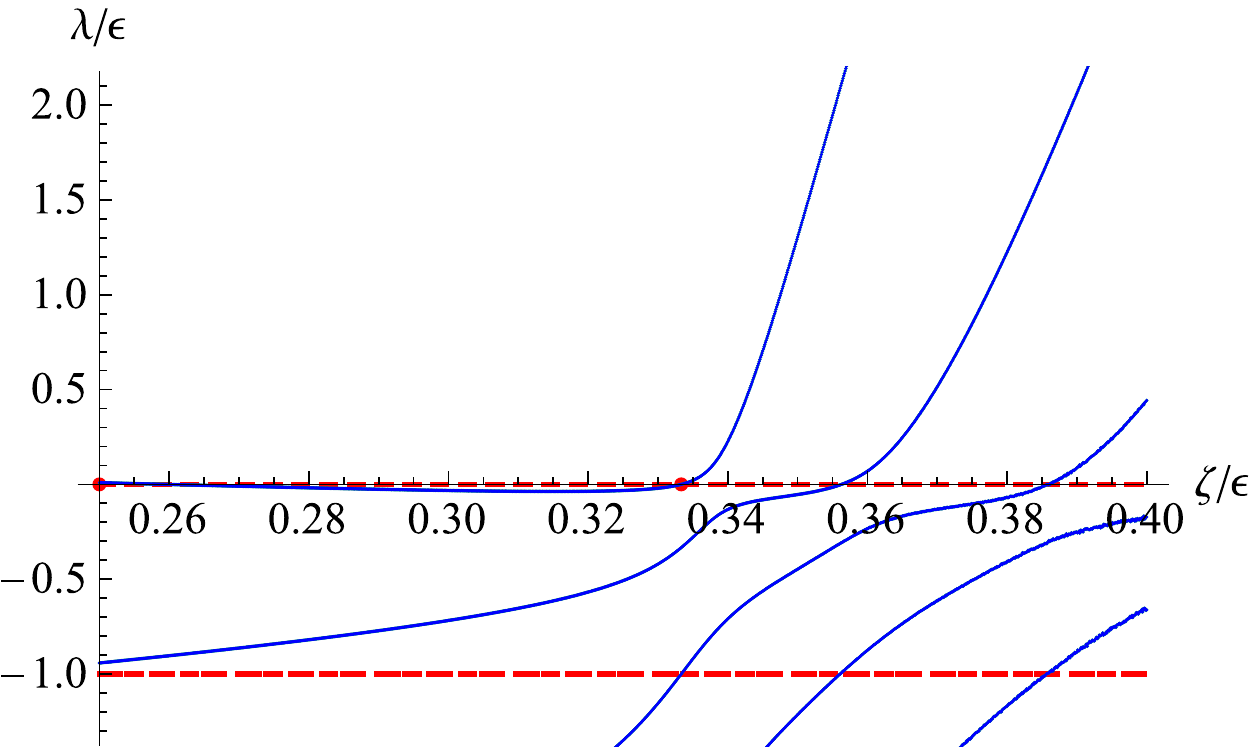}} \caption{Left: The real part $\mbox{Re}(\lambda)$ of the  eigen-values of Eq.~(\ref{31}) in the space of polynomials up to degree 15. Real  Eigen-values are shown in blue, complex pairs in green. Right: ibid, for $n=24$ for the five largest eigenvalues.  The red points mark $\zeta=\frac\epsilon4$ and $\zeta=\frac\epsilon3$. Note that the approximation for $R_\zeta(u)$ is exact for $\zeta=\frac\e3$ and $\zeta=\e$.  The eigenmodes $\lambda=0$ and $\lambda=-\epsilon$ are given only approximately. Indeed, one can check that among the eigenfunctions of those eigenvalues close to 0 and $-1$ there is one each which strongly resembles  the eigenfunctions  (\ref{38}) and (\ref{39}). We interpret the errors as an effective level-repulsion induced by the truncation scheme. This error reduces only very slowly when going to higher orders.}
\label{f:lambdaofzeta}
\end{figure*}
\section{Stability analysis}
\label{s:StabilityAnalysis}
\subsection{General setting}
Consider perturbations of Eq.~(\ref{flow-rescaled}) around a fixed point $R_\zeta$ with exponent $\zeta$, setting
$
R(u) = R_{\zeta}(u) + \eta f(u) + ...$.
Then, to linear order in $\eta$, the flow for the perturbation $f(u)$ is 
\beq \label{31}
- m \partial_m f(u) = (\epsilon-4 \zeta) f (u) + \zeta u f '(u) +  f''(u)  R''_\zeta(u)\ .
\eeq
In general, we look for eigen-modes $f_\lambda$ of the form 
\be
- m \partial_m f_\lambda(u) = \lambda f_\lambda(u)\ .
\ee
If we find $f_\lambda(u)$ with   $\Re(\lambda)>0$, then the fixed-point solution $R_\zeta(u)$ is linearly unstable. Absence of such a solution does not necessarily allow to conclude  linear stability; the stability matrix may, {\em e.g.}   be non-diagonalizable.

\subsection{Stability of the solution with $\zeta=\epsilon/3$}
We first establish the stability of  the {\em numerically chosen} solution $\zeta=\frac\epsilon 3$   w.r.t.\ higher-order polynomials. 
Setting $f(u) = (1-u)^n$, the r.h.s.\ of \Eq{31} becomes  
\bea
- m \partial_m (1-u)^n &=& \epsilon \bigg[ -\frac{ (n-2) (n-1)}{6} (1-u)^n \nn\\
&& ~~~~-\frac{(n-2) n}{3}  (1-u)^{n-1}\bigg] \ .
\eea
This proves that the highest-order term,  proportional to $(1-u)^n$ in the Taylor-expansion around $1$ decays, and a subdominant one proportional to $(1-u)^{n-1}$ is generated. Finally, all terms with Taylor-coefficients larger than $2$ are eliminated. 
Hence,  the solution with $\zeta=\frac\epsilon 3$ is linearly stable w.r.t.\ polynomial perturbations. 

\subsection{Two exact eigen-modes}

If $R_\zeta(u)$ is a solution of the flow-equation, then  
$
\eta^{4} R_{\zeta}(u/\eta)
$ also is a solution, implying that  there is a marginal eigen-mode  
\be\label{38}
f_{0}(u) = \partial_{\eta} \big|_{\eta=1} \eta^{4} R_{\zeta}(u/\eta) = 4 R_\zeta(u) -  u R_\zeta'(u)\ .
\ee
Another eigenvector exists for ${\lambda = -\epsilon}$, as can be verified directly\footnote{It can be constructed from a superposition of the RG-fixed point and the marginal eigen-vector, see section VII.B of \cite{LeDoussalWiese2003a}, where the twice derived form is given.}
\be\label{39}
f_{-\epsilon}(u) = (\epsilon-4 \zeta)R(u)+ \zeta u R'(u)\ .
\ee

\subsection{Stability analysis for generic values of $\zeta$}
Now consider $\zeta\neq \frac\epsilon3$.  The flow of a perturbation proportional to $(1{-}u)^n$ then becomes \bea
- m \partial_m (1{-}u)^n &=&  \frac{( \zeta {-}\e) (3  \zeta {-}2\e) (3 \zeta {-}\e) (n{-}1) n}{576  \zeta ^2} (1{-}u)^{n+2} \nn\\
& & + \frac{( \zeta {-}\e) (3  \zeta {-}\e) (1{-}n) n}{72  \zeta }  (1{-}u)^{n+1} \nn\\
&& +  \! \left( \frac{1}{4} ( \zeta {{-}}\e) (n{{-}}1) n{+} \zeta  n{+}\e {{-}}4  \zeta \right)(1{-}u)^n \nn\\
&& + (n{-}2)n  \zeta (1{-}u)^{n-1}\bigg]\ .
\eea
First note that the space spanned by $\{1,1-u \}$ has no non-linear terms, hence the rescaling terms lead to $\lambda_0 = \epsilon - 4 \zeta <0$, and $\lambda_{1} = \epsilon-3 \zeta$. The latter is positive for $\zeta<\frac\epsilon 3$.  It is however {\em questionable} whether these solutions are {\em allowed by the boundary conditions}. 

Higher-order terms are generated for all  values of $\zeta$ other than $\zeta=\frac\epsilon 3$ or $\zeta=\epsilon$, and these solutions are {\em potentially} unstable.  The solution $\zeta=\epsilon$ does not produce terms of order $(1-u)^{n+2}$ or $(1-u)^{n+1}$, but the coefficient of the term $(1-u)^{n}$ becomes $\epsilon(n-3)$, i.e.\ these terms grow for $n>3$, contrary to the solution $\zeta=\frac\epsilon 3$, for which they decay. We can thus conclude that $\zeta=\epsilon$ is {\em unstable}.

The situation for generic values of $\zeta$ is delicate, as we will see in a moment: The strategy we followed was to  
  construct  for fixed $\zeta$ a Taylor expansion of $R_\zeta(u)$ around $u_{\rm c}=1$ up to order $n$, starting at $n=2$ (the space $n\le 1$ discussed above decouples). We then solved the Eigenvalue problem (\ref{31}) in the space of polynomials up to degree $n$ in $(1-u)$. For $n=15$ the result is plotted on figure \ref{f:lambdaofzeta} (left), along with a blow-up for $n=24$ (right). (These are the maximal values of $n$ to not induce significant numerical errors in the chosen range of $\zeta$). 
As figure   \ref{f:lambdaofzeta} shows, solutions with $\zeta>\frac\e 3$   are unstable. The domain with $\frac\e 4 \le \zeta\le \frac\e 3$ is {\em seemingly}  stable. Our analysis is, however, bugged with a truncation problem that does not disappear for larger $n$: The exact eigenmodes $\lambda=0$ and $\lambda = -\epsilon$ can only be found approximatively in figure \ref{f:lambdaofzeta} (right), where they are indicated by   red dashed lines. We interpret these findings in the sense that there is a strong level-repulsion between the eigenvalues, induced by the truncation. One can indeed check that the exact eigenfunctions   (\ref{38}) and (\ref{39}) are close to eigenfunctions in the spectrum of eigen-perturbations, even though  their eigenvalues are   off.

\bigskip

\subsection{Perturbing the solution $\zeta=\frac\epsilon4$}
The solution $\zeta=\frac\epsilon4$ is special. It has a continuous spectrum of eigen-functions, given with $\bar\lambda:=\lambda/\epsilon$ by \begin{widetext}
\bea
f_\lambda(u) &=& \frac{\Gamma \left(\frac{7}{6}-\frac{1}{6} \sqrt{25-16 \bar\lambda }\right) \Gamma \left(\frac{1}{6}
   \sqrt{25-16 \bar\lambda }+\frac{7}{6}\right) \, _2F_1\left(-\frac{1}{6} \sqrt{25-16 \bar\lambda
   }-\frac{5}{6},\frac{1}{6} \sqrt{25-16 \bar\lambda }-\frac{5}{6};\frac{1}{3};u^{3/2}\right)}{\Gamma
   \left(\frac{1}{3}\right)} \nn\\
   && -\frac{u \Gamma \left(\frac{11}{6}-\frac{1}{6} \sqrt{25-16 \bar\lambda }\right)
   \Gamma \left(\frac{1}{6} \sqrt{25-16 \bar\lambda }+\frac{11}{6}\right) \, _2F_1\left(-\frac{1}{6}
   \sqrt{25-16 \bar\lambda }-\frac{1}{6},\frac{1}{6} \sqrt{25-16 \lambda
   }-\frac{1}{6};\frac{5}{3};u^{3/2}\right)}{\Gamma \left(\frac{5}{3}\right)}\ . \qquad
\eea\end{widetext}
A similar continuous spectrum is {\em seemingly} found when numerically solving the eigenvalue problem around a numerically obtained solution $R_{\zeta}(u)$. We suspect that this analysis is invalid, since we do not impose  appropriate boundary conditions at $u=0$.

\subsection{Numerical stability analysis}
Our numerical simulation seems to chose the fastest available  stable solution, in accordance to general ``wisdom'' of non-linear systems; however the latter has never been proven.

\begin{figure}[t]
\Fig{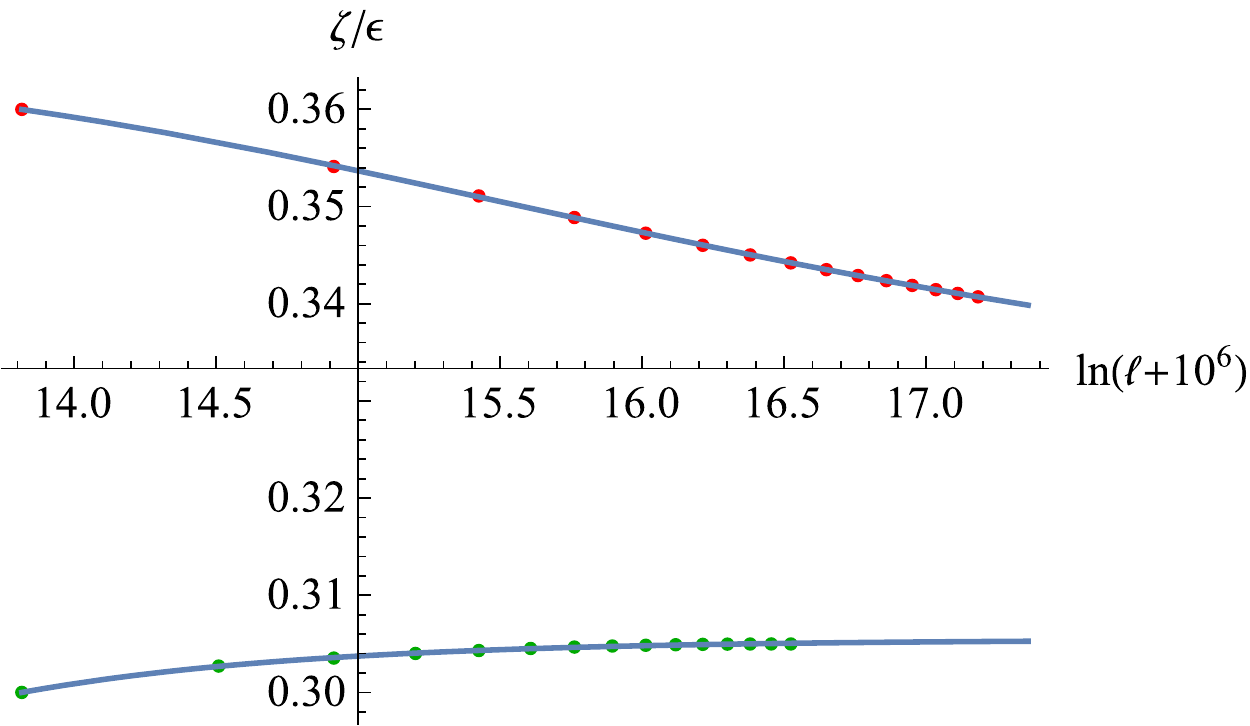}
\caption{Estimation of $\zeta$ from the relation $R_\ell(0) u_{\rm c}^{\frac{\epsilon}\zeta -4} \stackrel !=R_\zeta(0)$, starting from $R_\zeta(u)$ with $\zeta=0.36\epsilon $ (top) and $\zeta=0.3\epsilon$ (bottom). The two fits are $\zeta(\ell)/\epsilon = 0.325118\, +\frac{5.04594}{(l+2.78605\times 10^6)^{0.33522}}$ (top), and  $\zeta(\ell)/\epsilon = 0.305428\, -\frac{5561.08}{l+1.02487\times 10^6}$ (bottom).}
\label{f:convergence}
\end{figure}
We tried to start from another   fixed point in our family, and study whether it is stable when inserted into the flow equations and solved numerically. 
On figure \ref{f:convergence} we show how starting from fixed-point solutions with $\zeta=0.3 \e$ (bottom) and $\zeta=0.36\e$ (top) the RG flow evolves. 
While the solution with $\zeta>\frac\e 3$ seemingly converges towards the solution with $\zeta=\frac\e 3$, this is not the case for the solution with $\zeta<\frac\e 3$, in agreement with the stability analysis presented above. 
{\em Note however}, that even for $\zeta>\frac\e 3$ this convergence is on time scales that are orders of magnitude larger than the convergence from an analytic initial condition (see figure \ref{f:zeta-approach}).

\section{Higher-loop order}
\subsection{The flow equations}
At 2-loop order, we obtain the following functional renormalization group equation for the function \(R(u)\), 
\begin{eqnarray}
- m \partial_m   R(u) &=& (\epsilon-4\zeta) R(u) + \zeta u R'(u)+  \textstyle \frac{1}{2}
{R''(u)}^{2} \nn \\
 && +\left({\textstyle \frac{1}{2}} + \epsilon\, {\cal C}_{1} \right)
R''(u) {R'''(u)}^{2}     \nn \\
&& + {\cal C}_{2}   {R'''(u)}^4 + {\cal C}_{3}\, R''(u)  ^2 {R''''(u)}^2 \rule{0mm}{3ex} \nn \\ 
&&+ {\cal C}_{4}\,   R''(u) {R'''(u)}^{2}{R''''(u)}     \ ,\qquad \ \ \ 
\label{betafinal} 
\end{eqnarray}
with \begin{eqnarray}
{\cal C}_{1} & =&\frac{1}{4} +\frac{\pi ^{2}}{9}-  \frac{\psi'
(\frac{1}{3})}{6}= -0.3359768096723647...~~~ \label{C1}\\ \label{C2}
{\cal C}_{2} 
&=& \frac{3}{4}\zeta (3)+\frac{\pi ^{2}}{18}-\frac{\psi'
(\frac{1}{3})}{12}=0.6085542725335131... ~~~~~~~~\\ \label{C3}
{\cal C}_{3}&=&  \frac{\psi'
(\frac{1}{3})}{6} -\frac{\pi ^{2}}{9}= 0.5859768096723648... \\
{\cal C}_{4} \label{C4}
&=& 2+\frac{\pi ^{2}}{9}-\frac{\psi' (\frac{1}{3})}{6}= 1.4140231903276352... \ .
\end{eqnarray}
The first line contains the rescaling and 1-loop terms, the second line the 2-loop terms, and the last two lines the three 3-loop terms. The constants are related by \(\mathcal{C}_1=\frac14 -\mathcal{C}_3\), and \({\cal C}_4=2-{\cal C}_3 = \sqrt 2-0.000190372...  \)

\begin{figure}[t]
\Fig{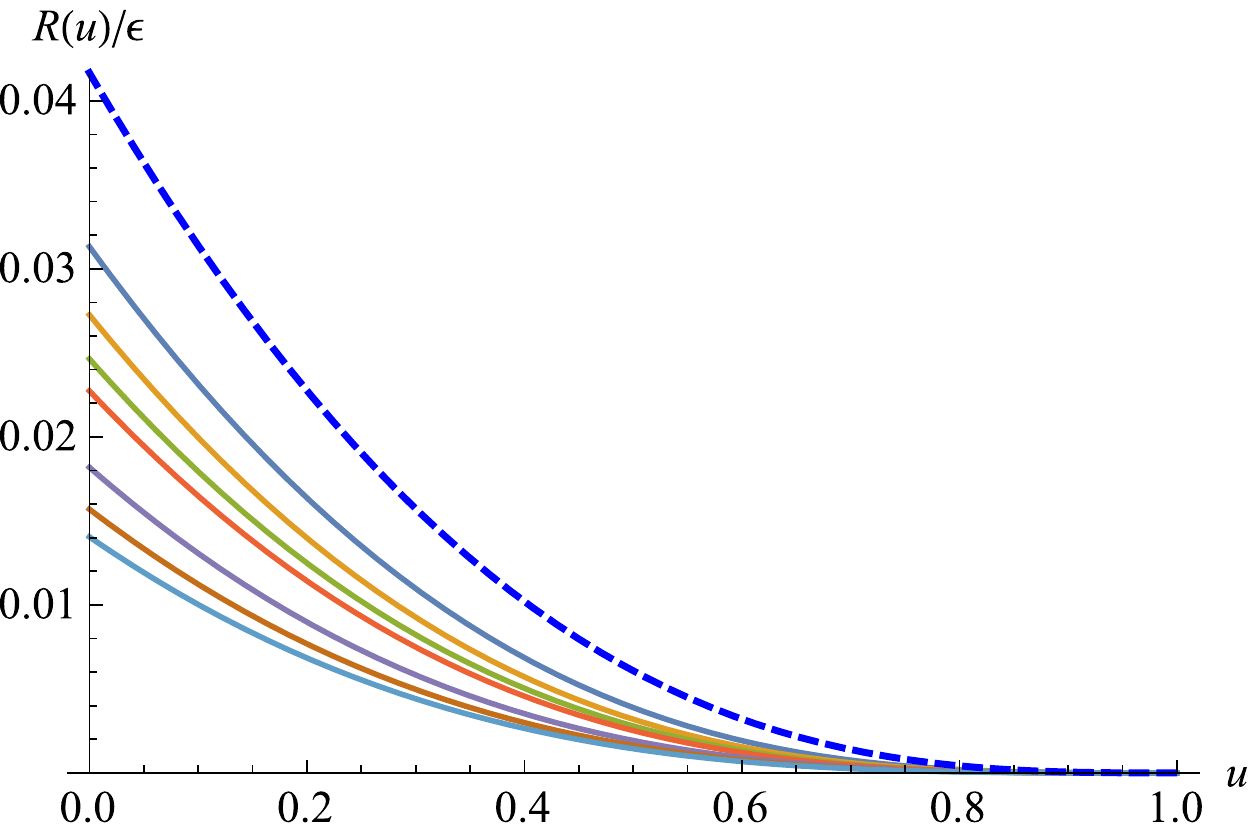} \caption{The 2-loop fixed point for $\zeta=\frac\e 3$; at 1-loop order (blue dashed line), and   2-loop order for $\epsilon=0.25,\, 0.5,\,0.75,\,1,\,2,\,3,\, 4$  (from top to bottom). The expansion is rather non-uniform.}
\label{f:2-loop}
\end{figure}

\subsection{Problems with the $\epsilon$ expansion, and a consistent series for the  2-loop fixed point}
Note that   at 2-loop order, we get a term 
\be
\frac12 R''(u) R'''(u)^2 \sim (1-u) + ...
\ee
close to the singularity when inserting any of our 1-loop solutions. Thus, there is no proper $\epsilon$-expansion for the fixed point at higher order. 

Valuable information is gained by studying the shock front generated by  the flow-equation at $n$-loop order. Without even calculating the Feynman diagrams, we   conclude that  
\beq
\partial _\ell R\big(u,u_{\rm c}(\ell)\big) \stackrel ! \simeq   \partial_u^{4 n}\left[  R\big(u,u_{\rm c}(\ell)\big) \right]^{n+1}\ .
\eeq
Supposing that 
\be
R\big(u,u_{\rm c}(\ell)\big) \simeq {\cal A} \left[ u_{\rm c}(\ell){-}u\right]^{\alpha}\ ,
\ee 
the latter equation implies for the shock front
\beq
  \left[ u_{\rm c}(\ell){-}u\right]^{\alpha-1} \partial_t u_{\rm c}(\ell) \stackrel ! \simeq  {\cal A}^{n}   \left[ u_{\rm c}(\ell){-}u\right]^{(n+1)\alpha-4 n}\ .
\eeq
This yields
\beq
\alpha = 4-\frac1n
\ .
\eeq
At 2-loop order, we should thus try an ansatz, starting at $(1-u)^{7/2} $ instead of $(1-u)^3$. 
Indeed, such an ansatz is possible, resulting into the series expansion
\bea
 {R(u)}  &=& \frac{16 \sqrt{\zeta } (1-u)^{7/2}}{105 \sqrt{5}} -\frac{2}{153} (1-u)^4 \nn\\
 &&  +\frac{4 (1-u)^{9/2}
   \left(2601 \sqrt{5} \zeta -5202 \sqrt{5} \epsilon +1540
   \sqrt{5}\right)}{15566985 \sqrt{\zeta }}  \nn\\
&&   -\frac{2 (1-u)^5 (-223686
   \zeta +200277 \epsilon -45800)}{1323193725 \zeta }\nn\\
   &&+ ... 
\eea
This series converges for all $0\le u \le 1$ as long as  $\epsilon \ge 0.2$. It is singular as a function of $\epsilon$. This is shown on Fig.\ \ref{f:2-loop}.

\begin{figure}[t]
\Fig{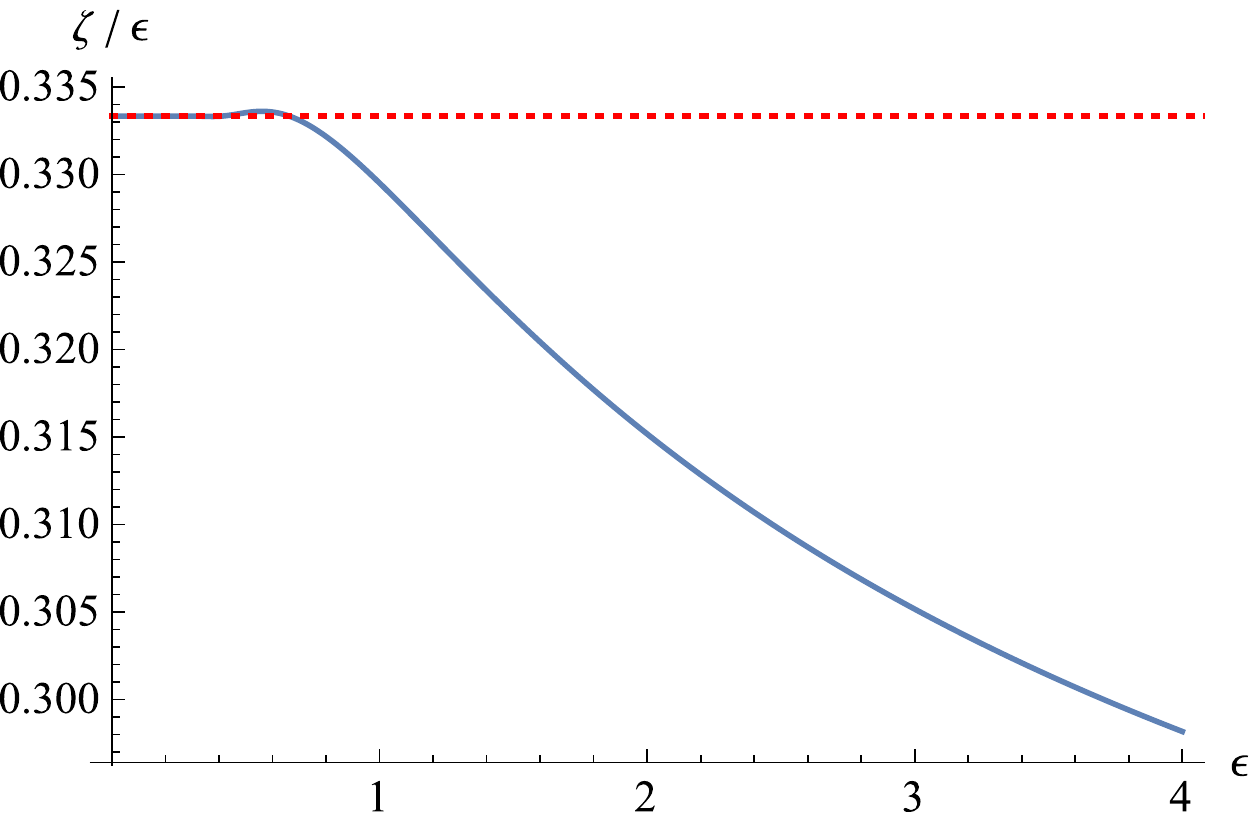}
\caption{The exponent $\zeta$ as obtained from the criterion $Q'(0) \stackrel != 0$, using the series expansion for the 2-loop fixed point up to order $(1-u)^{15}$. Points between $\epsilon=0$ and $\epsilon=0.25$ have been replaced by a straight line, which continues on the numerically obtained plateau of unknown origin up to about $\epsilon=0.4$.
}
\label{f:2-loop-zeta-from-slope}
\end{figure}

{\em If we use the same selection criterion as at 1-loop order}, namely that $Q'(0^+) = {R}'''(0^+) =0$, then this series expansion yields  the exponents presented on figure \ref{f:2-loop-zeta-from-slope}. Curiously, when $\epsilon \le 0.4$, the exponent $\zeta/\epsilon$ does not change; it then increases slightly, before decreasing for larger values of $\epsilon$.

\subsection{Numerical integration of the flow-equations}

We tried to confirm the above findings by a numerical integration of the flow-equations. Again we use  discrete derivatives, 
\bea
R''(u) &\simeq & \frac{ R(u{+}\delta u)+R(u{-}\delta u)-2 R(u) }{(\delta u)^2} \\
\!\!R'''(u) &\simeq & \frac{ R(u{+}2\delta u)-3 R(u{+}\delta u)+3 R(u) -R(u{+}\delta u)  }{(\delta u)^3} \ .\nn\\
\eea
Note that the third derivative involves two points to the right, and only one to the left. 
For the point $u=0$, we take (as at 1-loop order) the first 4 points except the very first ($i=2,3,4,5$) of the discrete derivative, and interpolate with a cubic polynomial, which is then extrapolated to $i=1$. 

\begin{figure}[t]
\Fig{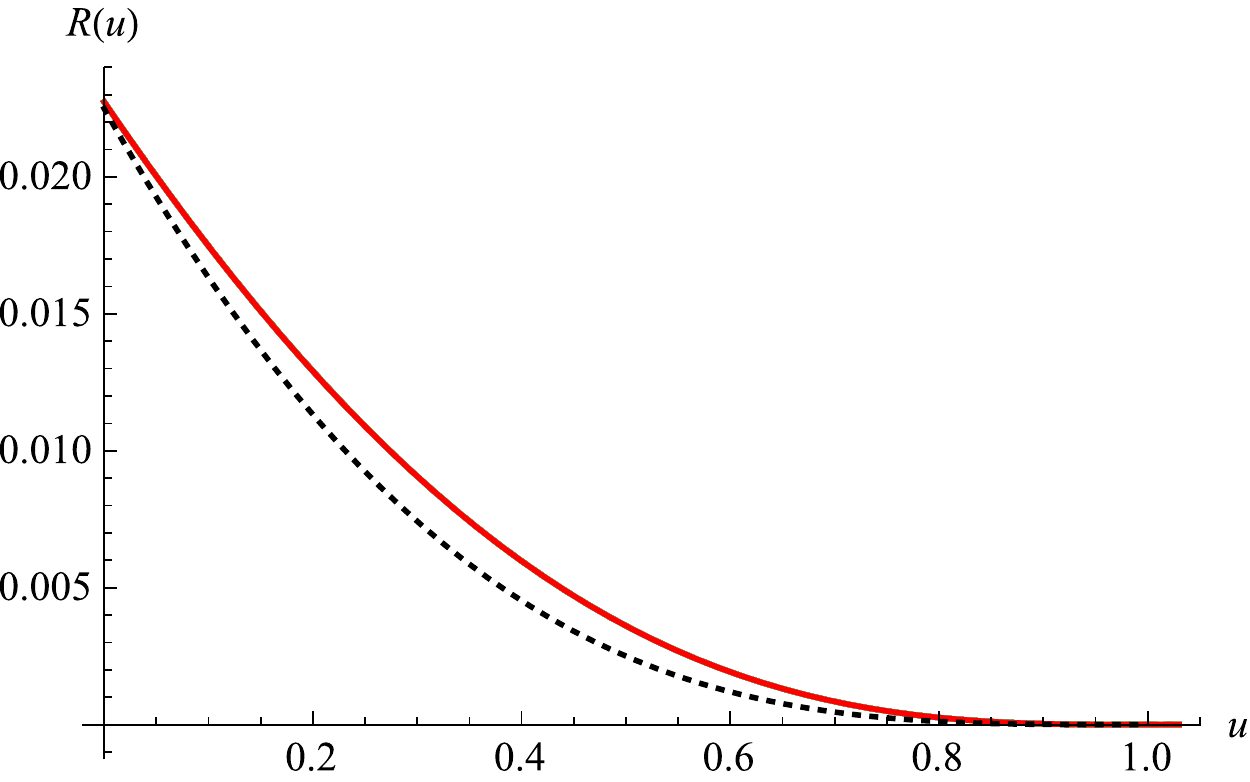}
\caption{The 2-loop fixed point $R(u)$ at $\epsilon=1$, as obtained from a numerical integration of the flow equations (solid red), compared to the 1-loop fixed point (dashed, black). Both solutions were rescaled to have $u_{\rm c}=1$, and $R(0)=1/ {24}$.}
\label{f:2-loop-num}
\end{figure}

Our initial condition is the 1-loop solution for $\zeta=\frac{\epsilon} 3$, at $\e=1$. We tried to use a smooth function to start with, but did not succeed to stabilize the system of equations. We have plotted the shape of the large-time solution on Fig.~\ref{f:2-loop-num}. It shows significant deviations from its 1-loop counterpart. We were not able to conclude on the asymptotic value of the  exponent $\zeta$.  
A stability analysis around the fixed-point solutions for $\epsilon=1$ yields results compatible with our 1-loop findings. 
 
\section{Related Models}
\subsection{Large $N$, and relation to the  KPZ   equation}
\label{s:largeN}
We now consider the generalizsation of our RG-equations to $N$ components, with the aim of taking the limit     $N\to \infty$. To this aim, we replace 
\be
\cR (u) \to  N \,\cR\!\left(\frac{\vec u^{2}}{2N}\right)
\ .
\ee
Taking into account the proper index structure,  we  arrive at the following (unrescaled) RG equations:
\bea
\partial_{\ell} \cR \!\left(\frac{\vec u^{2}}{2N}\right)  &=& \frac{N}2 \left[\nabla_{i}\nabla_{j}\, \cR\! \left(\frac{\vec u^{2}}{2N}\right)\right] ^{2}  \\
&=& \frac N2 \left[ \frac{\delta_{ij}}N  \cR'\!\left(\frac{\vec u^{2}}{2N}\right) + \frac{u_{i}u_{j}}{N^{2}} \cR''\!\left(\frac{\vec u^{2}}{2N}\right)\right] ^{2} \!\!\! .~~~~~~~ \nn\eea
Noting $x:= \frac{\vec u^{2}}{2N}$, this yields the flow equations for a generic number $N$ of components, 
\be
\partial_{\ell} \cR (x) = \frac1{2} \cR'(x)^{2} + \frac{2x}{N} \cR'(x) \cR''(x) + \frac{2 x^{2}}{N} \cR''(x)^{2}\ .
\ee
In the limit of $N\to \infty$, this reduces to 
\be\label{KPZ}
\partial_{\ell} \cR (x) = \frac1{2} \cR'(x)^{2} \ .
\ee
\begin{figure}[t]
\Fig{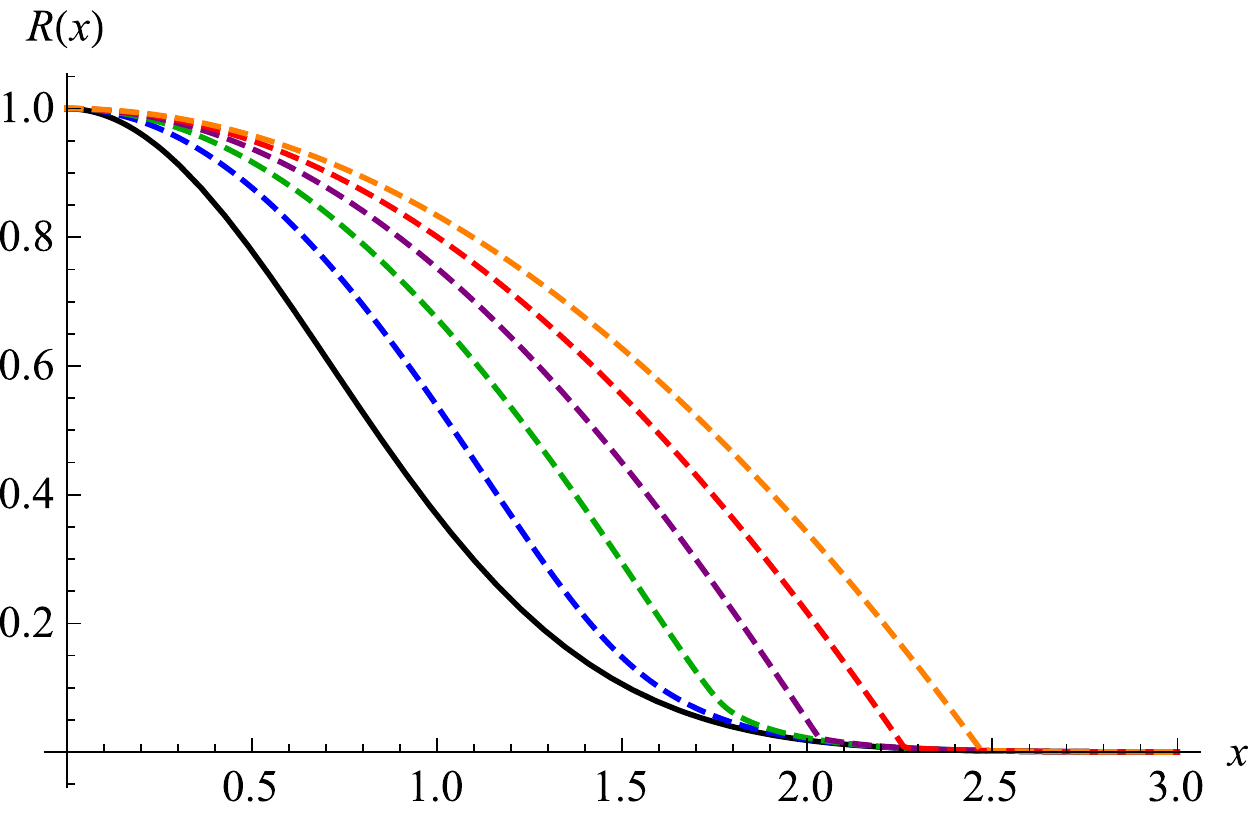}
\caption{Evolution of   Eq.~(\ref{KPZ}): Initial condition at  $\ell=0$ (black), followed by $\ell=1$ (blue, dashed), $\ell=2$ (green dashed),  $\ell=3$ (purple dashed) , $\ell=4$ (red dashed) and $\ell=5$ (orange, dashed).}
\label{f:KPZ}
\end{figure}This is the celebrated KPZ equation \cite{KPZ} in the limit of vanishing viscosity. 
Its solution is known analytically,  
\be\label{KPZ-explicit-solution}
\cR(x,\ell) = \max_{y} \left[  \cR(y,0) - \frac{2(x-y)^{2}}{\ell} \right]
\ .
\ee
As an example,  start with $\cR(x,0)= \rme^{{-x^{2}}}$. The result is shown on figure \ref{f:KPZ}. 
For large times the maxima are either    for $y=0$ or $y=x$, i.e.
\be\label{63}
\cR(x,\ell) \simeq \left\{ \begin{array}{cl} \displaystyle  1 - \frac{2 x^{2}}{\ell} & \mbox{ for } \displaystyle 0\le x\le\sqrt{\frac{\ell}2 } \\
 0 & \mbox{ for } \displaystyle x>\sqrt{\frac{\ell}2 }
 \rule{0mm}{5ex}
 \end{array}  \right. 
\ .
\ee
{\blue Note that $x^{2}\sim u^{4}$, yielding the standard $\phi^{4}$-fixed point in the domain $x<\sqrt{l/2}$.}

\begin{figure}[t]
\Fig{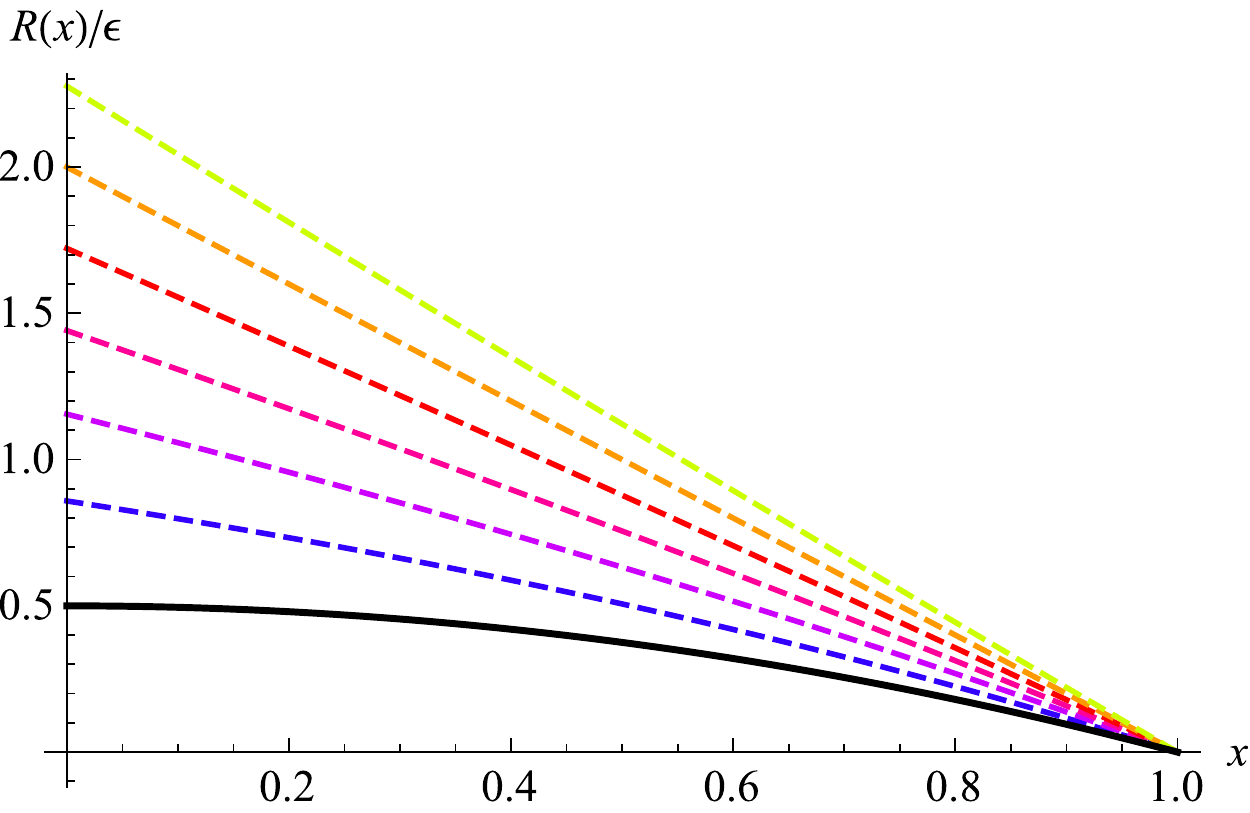}  \caption{Fixed points of Eq.~(\ref{large-N-rescaled}), for $\zeta=\frac\e4$ (black solid lines), and $\zeta=0.3 \e$, $0.35\e ... 0.6 \e$ (dashed lines, from bottom to top). Only the solution with $\zeta=\frac\e 4$ is stable.}
\label{f:sols-KPZ}
\end{figure}

The rescaled equation reads\footnote{Since $x\sim u^{2}$,   the rescaling term is $2\zeta x R'(x)$ instead of $\zeta u R'(u)$.}, 
\be\label{large-N-rescaled}
- m \partial_{m} R (x) = (\epsilon - 4\zeta) R(x)+ 2 \zeta x R'(x) + \frac1{2} R'(x)^{2} \ .
\ee
Since in our solution (\ref{63}) the function $\cR(0,\ell)$ is constant, this implies that $\epsilon - 4\zeta=0$, thus
\be
\zeta_{{{\rm large}~N}} = \frac{\epsilon}{4}\ .
\ee
There is a shock-front at $x_{c}= \sqrt{\ell/2}$, and the solution  $R(x,\ell)$ grows linearly for $x<x_{c}$, i.e.\ $R(x,\ell)\sim (x_{c}-x) \Theta(x\le x_{c})$. This is similar to what we observed in the preceding sections, but with a different exponent.  There is no ambiguity about the scaling. That does not mean that there are no solutions for other values of $\zeta$: Indeed, we found   a family of fixed points with $x_{c}=1$,  $R(x_{c})=0$, $R'(x_{c})=-4\zeta$, and series expansion 
\begin{figure}
\Fig{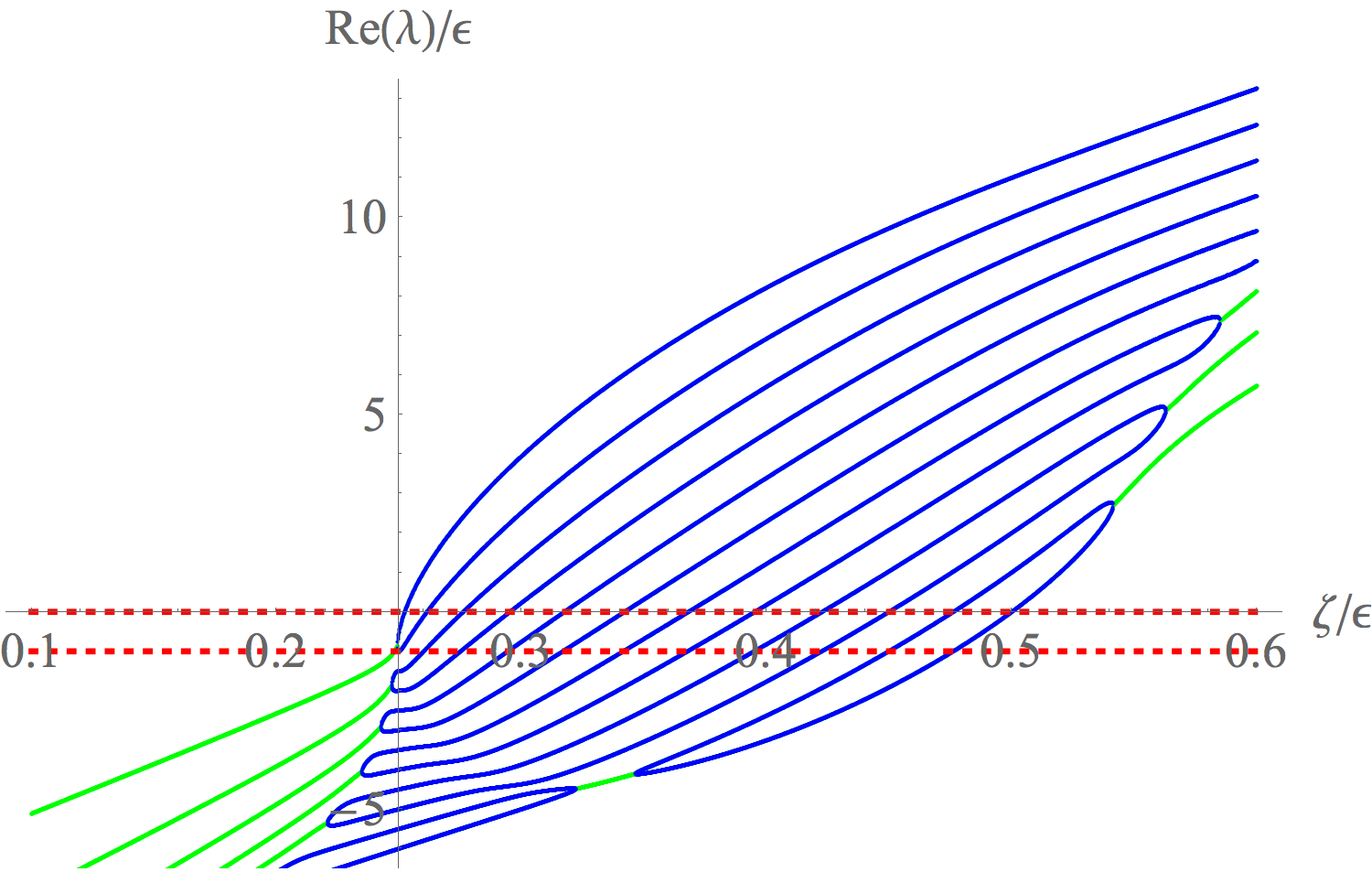}
\caption{Numerical stability analysis of the fixed-point solutions of Eq.~(\ref{large-N-rescaled}), at truncation order 12. All solutions with $\zeta>\frac\epsilon4$ are unstable. Real eigenvalues in blue, complex in green; the latter may be a result of the truncation. Several artefacts due to the truncation can be observed: First, the two exact eigenvalues $\lambda=0$ and $\lambda=\epsilon$ are not correctly reproduced. Second, the largest eigenvalue for $\zeta=\frac\epsilon4$ is not zero, but negative, leading to a small domain of stability for $\zeta/\epsilon$ up to 0.2523; this effect goes away for larger orders, for which however the rounding errors render the plot unreadable. We might only trust the information given for the largest eigenvalue. Finally, the domain for $\zeta<\epsilon/4$ does not contain a fixed point, as the function $R(x)$ is not defined down to $x=0$. This not-withstanding, we can study stability of the fixed point on a reduced domain for which the Taylor series would still converge; here the plot is instructive in telling us that $\zeta=\frac\epsilon4$ is at the boundary of an island of stability.}
\label{f:KPZ-FP-stability}
\end{figure}
\bea\label{KPZ-series}
R(x) &=&  -4 {  \zeta}  (x-1)+(2 {  \zeta} -\e) (x-1)^2 \nn\\
&& +\frac{1}{6} \left(8 {  \zeta}
   +\frac{\e^2}{{  \zeta} }-6\e \right) (x-1)^3 \nn\\
   && +\frac{(2 {  \zeta} -\e) (4 {  \zeta} -\e)
   (14 {  \zeta} -5 \e) (x-1)^4}{48 {  \zeta} ^2} \nn\\
   && +O(x-1)^5 \Big] 
\ .
\eea
As for the earlier fixed-point equation (\ref{FP-eq}), the series (\ref{KPZ-series}) converges only for 
\be\label{condition2}
\zeta_{{\rm large}~N} \ge \frac\epsilon4\ .
\ee
Integrating the fixed-point equation (\ref{large-N-rescaled}) numerically, starting at $x\approx 1$, we also find that only when condition (\ref{condition2}) is fulfilled, a solution exists from $x=1$ down to $x=0$. These solutions are shown on Fig.\ \ref{f:sols-KPZ}.

All solutions for $\zeta> \frac\e 4$ are unstable, and   flow to the fixed point with $\zeta=\frac\e 4$, as can bee seen from the explicit solution (\ref{KPZ-explicit-solution}). The latter conclusion is confirmed from a numerical stability analysis, presented on Fig.~\ref{f:KPZ-FP-stability}. Its flaws are discussed in the legend of the figure; they may serve as a guideline for  the stability analysis  presented in section \ref{s:StabilityAnalysis}, and which conclusions are robust, and which should be discarded. 

\subsection{Disordered elastic manifolds}
\begin{figure}[t]
\Fig{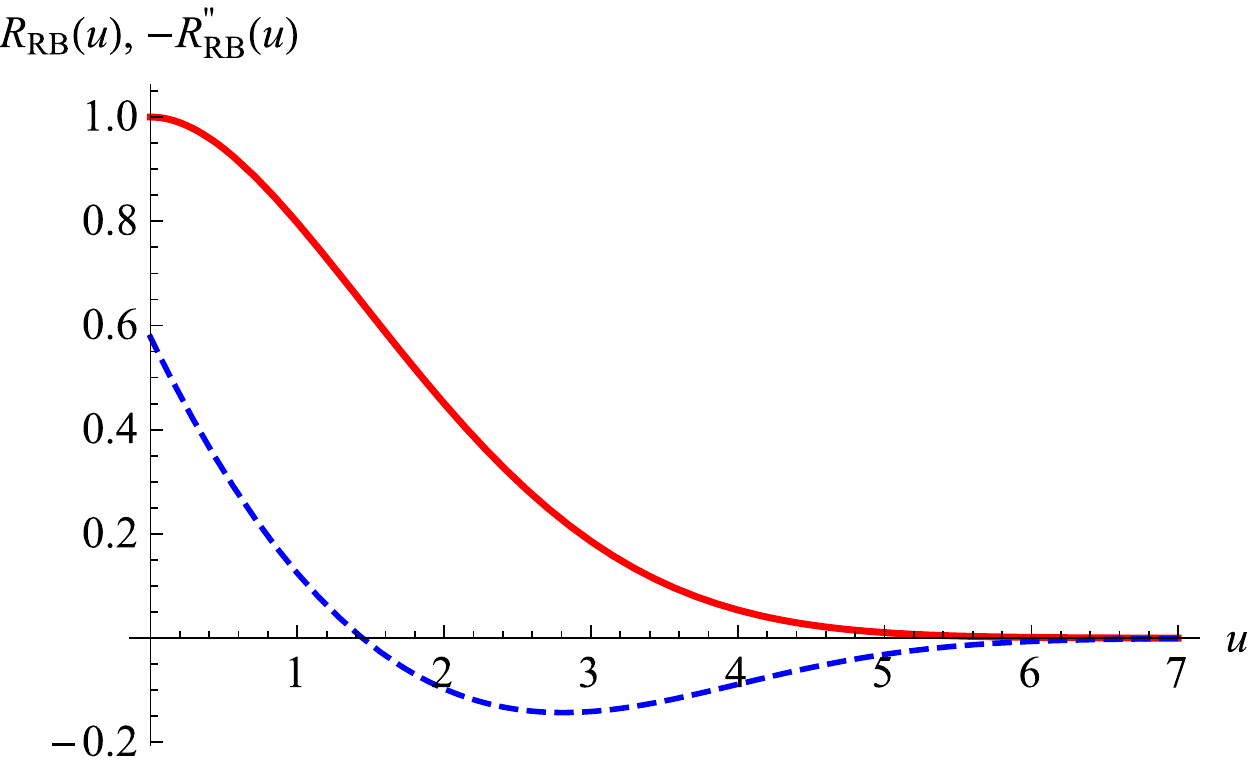}
\caption{The fixed-point function $R_{{\rm RB}}(u)$, and its second derivative $-R''_{{\rm RB}}(u)$ for the fixed-point equation (\ref{FP-disorder}). This fixed point is relevant for disordered elastic manifolds with random-bond disorder.}\label{f:RB}
\end{figure}

For disordered elastic manifolds, the central object of interest is the  disorder correlator $\cR$, defined as the disorder   average    of the potential-potential correlation function  (for an introduction see \cite{WieseLeDoussal2006}),
\be
\overline{ V(x,u ) V(x',u')}  = \cR(u-u') \delta^{d } (x-x')\ .
\ee
This correlator $\cR$ has at 1-loop order an RG-equation very similar to 
 Eq.~(\ref{flow-R-ell}) \cite{DSFisher1986,ChauveLeDoussalWiese2000a,LeDoussalWieseChauve2003}
\be\label{flow-R-ell-disorder}
\partial_\ell {\cal R}(u) =\half {\cal R}''(u)^2  -\cR''(u) \cR''(0) \ .
\ee
The presence  of the   term $ -\cR''(u) \cR''(0) $ is crucial. For the fixed point the equation to be solved is
\bea\label{FP-disorder}
- m \partial_m   R(u) &=& (\epsilon-4\zeta) R(u) + \zeta u R'(u) \nn\\
&& +   \frac{1}{2}
{R''(u)}^{2} -R''(u) R''(0)\ .
\eea
Starting from a Gaussian initial condition for $\cR(u)$, Eq.~(\ref{flow-R-ell-disorder}) flows to a fixed point
$
R_{{\rm RB}}(u)
$, solution of Eq.~(\ref{FP-disorder}), and displayed on Fig.~\ref{f:RB}, with 
\be
\zeta_{{\rm RB}} \simeq 0.2082981  \epsilon + ...
\ .
\ee
This fixed point is relevant for disordered elastic manifolds subject to random-bond, i.e.\ short-ranged  disorder.

The fixed point has the following properties \cite{ChauveLeDoussalWiese2000a,LeDoussalWieseChauve2003}: 
\begin{itemize}
\item[(i)] $R_{\rm RB}(0)>0$,  $R_{\rm RB}'(0^{+})=0$, $-R_{\rm RB}''(0^{+})>0$, $R_{\rm RB}'''(0^{+})>0$. 
Thus $R_{\rm RB}(u)$ is non-analytic at $u=0$, with a non-analyticity starting at order $|u|^{3}$. 

\item[(ii)] The fixed point has a Gaussian tail, i.e.\ there exists a constant $c$, s.t. 
$ R_{\rm RB}(u) < \rme^{{-c u^{2}}}$.

\item[(iii)] The solution is unique, and attractive. The largest two eigenvalues are $\lambda=0$ and $\lambda=-\epsilon$. 

\end{itemize}
Thus this fixed point has a singularity at $u=0$, {\em and  no singularity} at finite $u=u_{\rm c}<\infty$. 
It can be identified and measured in a numerical simulation \cite{MiddletonLeDoussalWiese2006}.
Let us also mention that  for elastic manifolds driven through a disordered environment, a different fixed point is relevant with $\zeta=\frac\epsilon3$. It can be measured in numerical simulations \cite{RossoLeDoussalWiese2006a}, and  in  experiments \cite{LeDoussalWieseMoulinetRolley2009}.

\section{Conclusions}
In this article, we considered the seemingly familiar setting of scalar field theories; the difference being that contrary to the standard $\phi^{4}$-potential ours is bounded, and quickly decays to 0. We found that under RG the effective potential develops a cusp at the origin, and a cubic singularity $\sim (u_{\rm c}-u)^{3}\Theta(u\le u_{\rm c})$  at a  shock front $u_{\rm c}$ increasing under RG. While there is an infinity of such solutions, our evidence suggests that a specific one is chosen {\em dynamically}, when starting from generic smooth initial conditions; this solution has a roughness exponent (dimension of the field) $\zeta = \frac\epsilon 3$. 

To put our findings into context, we discussed two similar equations: The first is as above, but at large number of components $N$, which maps onto the KPZ equation. Its analysis is much easier, leading to a roughness exponent $\zeta=\frac\e4$. The solution also contains a   shock-front, with a linear instead of a cubic singularity. 

In contrast, for disordered elastic manifolds, a similar flow equation pertains to the renormalization of the disorder correlator. For short-ranged initial conditions, it has a single fixed-point solution with well-defined roughness exponent $\zeta \simeq 0.2083 \e$, to which the flow naturally tends. It does not develop a shock singularity at finite $u_{\rm c}< \infty$. 

We finally solved a toy model presented in appendix \ref{a:toy}. Here, singularities appear only after  passing via a Legendre transform from the potential ${\cal W} = \ln {\cal Z}$ to the action $\Gamma$. Since at 1-loop order both objects   have the same RG-equation, we do not know how to interpret our findings.

Several problems remain a challenge for future research: 

\begin{enumerate}
\item[(i)]
What is the proper regularization of the  RG-equation?
\item[(ii)] Are there regularizations which lead to distinct critical exponents?
\item[(iii)] What is the physical interpretation of this new fixed point? Is it relevant for wetting?
\item[(iv)] What is the phyical interpretation of the cusp, and of the cubic shock front?

\item[(v)] What is the proper protocol for a simulation?

\end{enumerate}

\begin{figure}[t]
\vspace*{-3mm}
\Fig{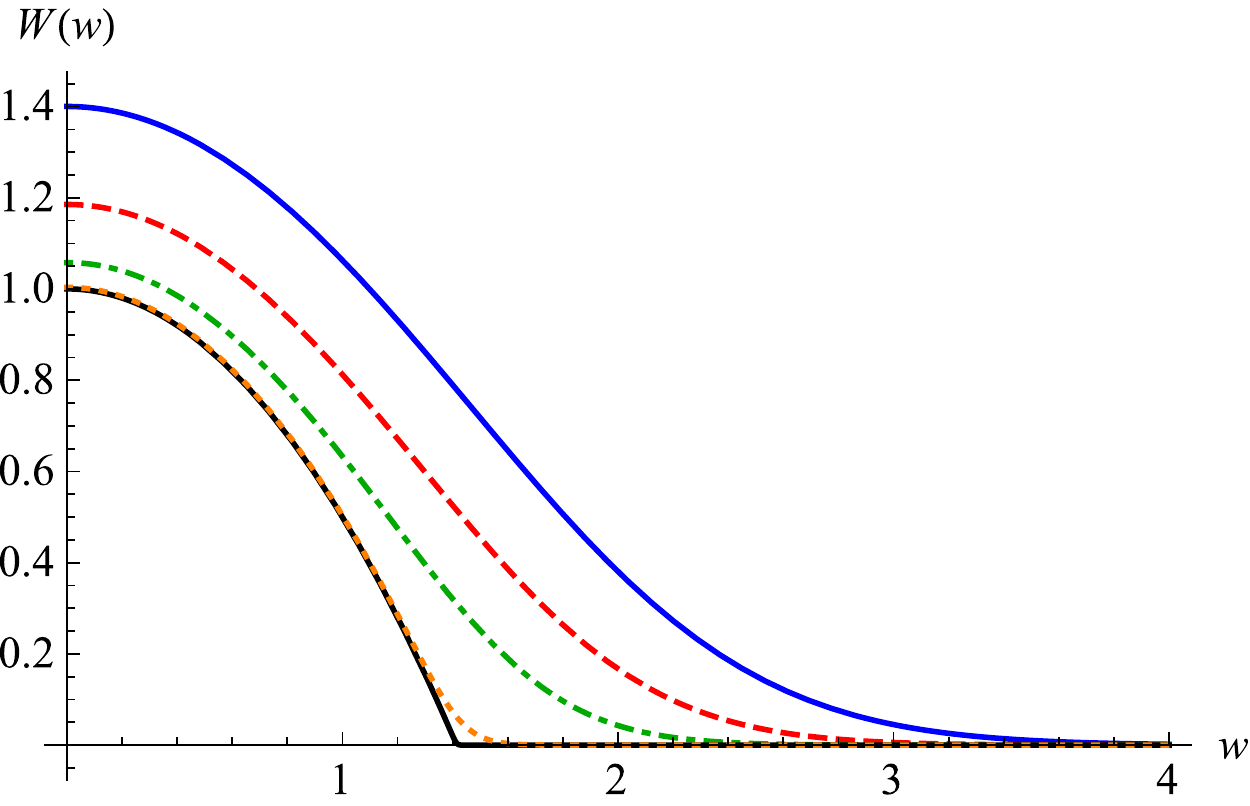}
\caption{$W(w)$ for the toy model (\ref{toy})~ff.~with  $\alpha=-1$.  From top to bottom: $m=0.4$ (blue solid), $m=0.25$ (red dashed), $ m=0.1$ (green, dot-dashed), $m=10^{-5}$ (orange, dotted), and the limit of $m\to 0$ (black, solid). Both axes are rescaled so that the limit of $m\to 0$ exists. (This resealing is a factor of $1/\ln(1/m)$ for the $W$-axes, and a factor of $1/\sqrt{\ln(1/m)}$ for the $w$-axis.)  }
\label{f:Wofw}
\end{figure}

\acknowledgements
It is a pleasure to thank Eduoard Br\'ezin  for enlightening discussions, and for asking all the right questions.
Many thanks   go to Fran\c cois David and Andrei Fedorenko for patiently  listening and pointing out weak points.  
I am   grateful to Martine Ben Amar,  Bernard Derrida and Vincent Hakim for sharing their insights into the   analysis of non-linear systems. 

Financial 
 support from PSL through  grant No.\ ANR-10-IDEX-0001-02-PSL is gratefully acknowledged.

\appendix

\section{Single degree of freedom: A toy model}
\label{a:toy}
We consider the following toy model:
\bea \label{toy}
V_{0}(u) &=& - \rme^{{-u^{2}}} \\
V_{m}(u,w) &=& \frac{m^{2}}2 (u-w)^{2}
\ .
\eea
We define 
\be \label{74}
W_{\rm toy}(w):=  \ln \!\left( \frac {m^{\alpha}} {\sqrt{2\pi}}\int_{-\infty}^{\infty} \rmd u\,\rme^{- V_{0}(u)- V_{m}(u,w)}\right)
\ .
\ee
We can rewrite Eq.~(\ref{74}) as 
\be\label{73}
 W_{\rm toy}(w)= \ln \!\left( 1{+}\frac {m^{\alpha}} {\sqrt{2\pi}}\int_{-\infty}^{\infty} \rmd u \left(\rme^{- V_{0}(u)}{-}1 \right) \rme^{- V_{m}(u,w)}\right) .
\ee
Note that we added a factor of $m^{\alpha}$ to mimic for the factor of $m^{2-d}$ present in Eq.~(\ref{basic}).
Formula (\ref{73}) can more precisely be evaluated numerically, since the integral can be cut off at, say $u = \pm 10$. This implies the asymptotic behaviour for small $m$, 
\bea
 W_{\rm toy}(w) &\simeq&  \ln \left( 1+\frac {m^{\alpha}} {\sqrt{2\pi}}  \rme^{- V_{m}(0,w)} \int_{-\infty}^\infty   (\rme^{\rme^{-u^2}}-1)\, \rmd u \right)\nn\\
 &=&   \ln \left( 1+1.043\,   {m^{\alpha}} \,  \rme^{- V_{m}(0,w)}  \right) \nn\\
 &\approx&  1.043\,   {m^{\alpha}} \,  \rme^{- V_{m}(0,w)} - 0.5528  {m^{2\alpha}} \,  \rme^{-2 V_{m}(0,w)} \nn\\
 && +...   \label{76}
\ .
\eea 
If $\alpha$ is positive, 
 the linear term dominates for small $m$, corresponding  to the first term in Eq.~(\ref{basic}); a linear approximation of the flow equations is appropriate.  On the other hand, if $\alpha<0$, the limit is non-trivial, and $W$ develops a cusp for $m\to0$. This is presented on figure \ref{f:Wofw}. 

\begin{figure}[t!]
\vspace*{-3mm}
\Fig{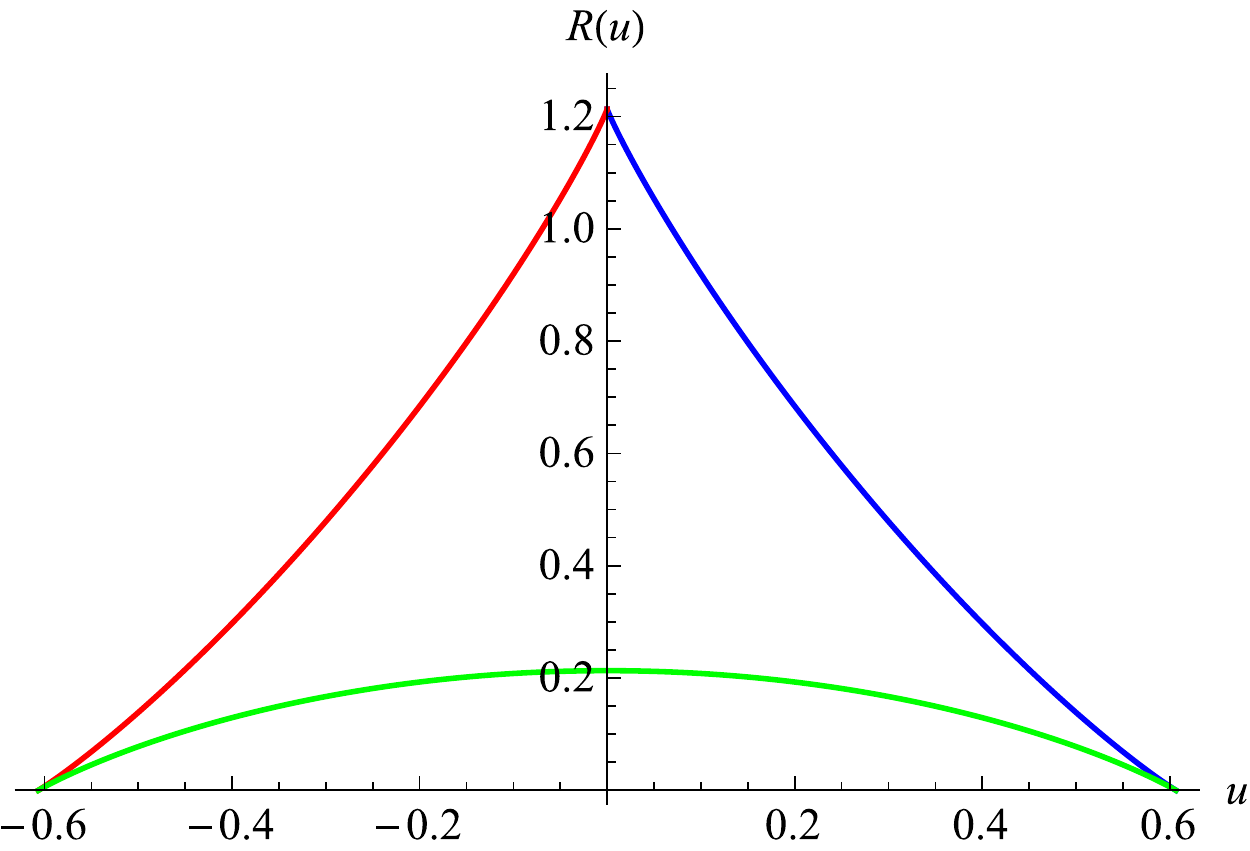}
\caption{$R(u)$ for the toy model (\ref{toy})~ff.~with $m=1$, starting from Eq.~(\ref{76}), and using (\ref{A7}). }
\label{f:R-particle2}
\end{figure}

On the other hand, for large $m$ and $\alpha=1$, we recover the initial condition, 
\be
 W_{\rm toy}(w)  \simeq -  V_0(w)\ .
\ee
We confirmed both limits numerically. Let us now perform the Legendre transform (for  $m=1$)
\be\label{A7}
W_{\rm toy}(w) + R_{\rm toy}(u) -R_{\rm toy}(0) = - u w \ .
\ee
We have added a constant $R(0)$, s.t.\ we can put $R_{\rm toy}(\infty) \to 0$. 
This transformation is most easily performed numerically, plotting $u(w) = -W_{\rm toy}'(w)$ versus $R_{\rm toy}(u(w)) -R_{\rm toy}(0)= W_{\rm toy}'(w) w- W_{\rm toy}(w)$.  Graphically, this amounts to drawing the tangent to $W_{\rm toy}(w)$, and   tracking the intersection of this tangent with the vertical axis.  The outcome of this construction   is shown on figure \ref{f:R-particle2}. It has three branches: $R_{\rm toy}(u)$ starts with a linear slope  at $u=0$, resulting from $W_{\rm toy}(w)$ with $w>1$ (blue curve). It terminates at $u_{\rm c}>0$, with a term proportional to $(u-u_{\rm c}) \theta(u<u_{\rm c}) $. The red branch is from $w<-1$. The green branch is the image of $-1<w<1$. 

It remains to be clarified what this toy model   teaches us  about the problem at hand; especially, shall we compare $W$ or $R$ with the results discussed in the main text?

\pagebreak[5]

\tableofcontents

\end{document}